\documentclass{article}
\pdfoutput=1
\usepackage{lscape}
\usepackage{lipsum}
\usepackage{enumerate}
\usepackage{amsfonts}
\usepackage{amsmath}
\usepackage{comment}
\usepackage{textcomp}
\usepackage{amsthm, tikz}
\usepackage{amssymb}
\usepackage{color}
\usepackage{graphicx}
\usepackage{pdflscape}
\usepackage{afterpage}
\usepackage[matrix,arrow,curve]{xy}
\usepackage{array}
\usetikzlibrary{positioning}
\usetikzlibrary{arrows}

\usepackage[ref,super]{cite}

\usepackage{braids}

\makeatletter
\renewcommand*{\@cite@ofmt}{\bfseries\hbox}
\makeatother

\hoffset= - 1.2in         

\begin{document}

\title{\vspace{0.1cm}{\Large {\bf Eigenvalue hypothesis for multi-strand braids}\vspace{.2cm}}
\author{{\bf Saswati Dhara$^{a}$},\ {\bf A. Mironov$^{b,c,d}$}, \ {\bf A. Morozov$^{b,c,d}$}, \ {\bf An. Morozov$^{c,d}$}, \\   {\bf P. Ramadevi$^{a}$}, \ {\bf Vivek  Kumar Singh$^{a}$},\
{\bf A. Sleptsov$^{c,d,e}$}}
\date{ }
}

\maketitle

\vspace{-5.5cm}

\begin{center}
\hfill FIAN/TD-24/17\\
\hfill IITP/TH-18/17\\
\hfill ITEP/TH-31/17\\
\end{center}

\vspace{4.2cm}

\begin{center}

$^a$ {\small {\it Department of Physics, Indian Institute of Technology Bombay, Mumbai 400076, India}}\\
$^b$ {\small {\it Lebedev Physics Institute, Moscow 119991, Russia}}\\
$^c$ {\small {\it ITEP, Moscow 117218, Russia}}\\
$^d$ {\small {\it Institute for Information Transmission Problems, Moscow 127994, Russia}}\\
$^e$ {\small {\it Laboratory of Quantum Topology, Chelyabinsk State University, Chelyabinsk 454001, Russia }}
\end{center}

\vspace{1cm}

\begin{abstract}
Computing  polynomial form of the colored HOMFLY-PT for non-arborescent knots obtained from three or more strand braids is still an open problem. One of the efficient methods suggested for the three-strand braids relies on the eigenvalue hypothesis which uses the Yang-Baxter equation to express the answer through the eigenvalues of the ${\cal R}$-matrix. In this paper, we generalize the hypothesis to higher number of strands in the braid where commuting relations of  non-neighbouring $\mathcal{R}$ matrices are also incorporated.  By solving these equations,  we determine the explicit form for $\mathcal{R}$-matrices and the inclusive Racah matrices in terms of braiding eigenvalues (for matrices  of size up to 6 by 6). For comparison, we briefly discuss the highest weight method for four-strand braids carrying fundamental  and symmetric rank two $SU_q(N)$ representation. Specifically, we present all the inclusive Racah matrices for representation $[2]$ and compare with the matrices obtained from eigenvalue hypothesis.

\end{abstract}


\vspace{.5cm}

\tableofcontents

\section{Introduction}
\setcounter{footnote}{1}

Classification of knots is one of the challenging research problem.   The well
known Jones', HOMFLY-PT and Kauffman polynomials \cite{knotpols,Con} can distinguish many inequivalent knots but not all knots. Witten's pioneering work\cite{wit} involving Chern-Simons field theory \cite{CS} and Jones' polynomials suggested that the generalized knot invariants can be computed for any knot ${\cal K}$ carrying arbitrary representation $R$ of any gauge group ${\cal G}.$  They are referred to as  colored knot invariants $H_R^{\cal K}$ which are supposed to give a pool of data to  attempt the famous challenging problem of `classification of knots'.

The methodology of writing the knot invariants is straightforward and they involve braiding
eigenvalues and  Racah matrices. However, the polynomial form of such invariants can be determined only if we know the Racah matrices. In fact,
the Racah matrices are fully known \cite{su2Racah} only for $SU_q(2)$ enabling the evaluation of colored Jones polynomial for any knot. Recently, from colored HOMFLY-PT for twist knots \cite{satoshi} and pretzel knots \cite{MMSRacah} (using the evolution method \cite{DMMSS,db,evo} in the latter case), closed form expression for Racah matrices  involving
any $SU_q(N)$ symmetric representation was conjectured \cite{nrzRacah,MMSRacah,MMShyper}. Subsequently colored HOMFLY-PT for arborescent knots \cite{Con,arbor} carrying symmetric representations were computable \cite{IMMMfe,satoshi,arborcalc,MMSRacah}.
Further, colored HOMFLY-PT for some rectangular\cite{rect}  and non-rectangular representations \cite{non-rect, Rama} of $SU_q(N)$ are obtained for arborescent knots. It is still an open problem to compute colored HOMFLY-PT for any non-arborescent knot carrying symmetric and other representations.

Based on  Reshetikhin-Turaev  (RT) approach\cite{RT} and their variants, new methods have been devised to obtain colored invariants for non-arborescent knots from closure
of three or more strand braids. In \cite{AMMMpath} a universal construction for knots in the fundamental representation was suggested but has not been generalized to higher representations. Another approach, used in Refs.\cite{MMMII,IMMM3,MMMS21,MMMS31,SSh},
involves calculations of the highest weight vector for various representations which becomes computationally tedious as we increase the number of braids.  We refer the reader to
see the papers \cite{finger,Rama2} where some non-arborescent knot invariants are presented from generalisation of the method\cite{Rama}.
Even though, these approaches are straightforward, the evaluation process becomes cumbersome. Another powerful method called eigenvalue hypothesis was  suggested in Ref. \cite{IMMMec}. In fact, we will focus on the essential details of the eigenvalue hypothesis  which will provide some inclusive Racah matrices to simplify the tedious highest weight approach.

\bigskip

\textbf{The eigenvalues hypothesis claims that the inclusive Racah matrix is fully determined by the eigenvalues of the $\mathcal{R}$-matrix.}

\bigskip
Using the suggested hypothesis for the three-strand case \cite{IMMMec}, the inclusive Racah matrices up to size $5 \times 5$  in terms of the $\mathcal{R}$-matrix eigenvalues have been guessed for
$SU_q(2)$
\footnote{It works in the following way: let us fix $V$ and $Q$ to be the spin $j$ and $3j-k+1$ representations of $SU_q(2)$ respectively. Then, the inclusive Racah matrix $U_{k\times k}(j)$ for $V^{\otimes 3}\to Q$ has size $k\times k$, and the $k$ eigenvalues of the ${\cal R}$-matrix, $\lambda_i(j)$, $i=1,\ldots,k$ are parameterized by $j$.} and later were shown to work for arbitrary $SU_q(N)$.  Subsequently,  $6\times 6$ inclusive Racah matrix was calculated in terms of the eigenvalues \cite{Univ}, using the V\"ogel universality hypothesis of Chern-Simons theory\cite{CSuniv,MMMkr}. These conjectured matrices have been independently checked\cite{ecA}.

Let us briefly discuss the construction of inclusive Racah matrix\cite{MMMII} for three-strand braid. This  will set the stage to go to four or more strand braids.
For three-strand braid carrying $SU_q(N)$ representations $V_1,V_2,V_3$, it is well-known that the Racah matrix $U_{ij}$ (originally introduced by G. Racah and E.P. Wigner) is the matrix that intertwines the maps
$$
(V_1\otimes V_2)\otimes V_3  \longrightarrow Q \qquad \text{and} \qquad V_1\otimes (V_2\otimes V_3) \longrightarrow Q
$$
The matrix indices of $U_{ij}$ are enumerated by the representations from $V_1\otimes V_2=\oplus_i T_i$ (the first index) and from $V_2\otimes V_3=\oplus_j T_j$ (the second index).
Since we are interested in the construction of knots from braids, we need to consider $V_1=V_2=V_3=V$. In fact $U_{ij}$ are the
matrices that relate the braiding matrix (${\cal R}_1$) acting on the first two strands and the braiding matrix (${\cal R}_2$) on the second and third strands in the three-strand braid.
Note that $U_{ij}$ are referred as inclusive Racah matrix when $V \neq Q$  and exclusive Racah matrix when $V=Q$.

Whatever the representations $V$ and $Q$ are, if the Racah matrix has size $k\times k$, it is expressed through $k$ eigenvalues of the $\mathcal{R}$-matrix ${\cal R}:\ V\otimes V\to V\otimes V$. These eigenvalues are very simple: $\pm q^{C_2(X)}$, where $C_2(X)$ is the eigenvalue of the second Casimir operator in representation $X\in V\otimes V$.
The evaluation of $U_{ij}$ is governed by the Yang-Baxter equation
\begin{equation}\label{YB}
\boxed{
\mathcal{R}_1\mathcal{R}_2\mathcal{R}_1=\mathcal{R}_2\mathcal{R}_1\mathcal{R}_2.}
\end{equation}
Particularly, one can diagonalize $\mathcal{R}_1$ and take $\mathcal{R}_2=U\mathcal{R}_1U^\dagger$. Then (\ref{YB}) relates the Racah matrix $U$ to the eigenvalues of ${\cal R}_1$.
In principle,  for various diagonal $k \times k$ $\mathcal{R}_1$, the elements of $k \times k$ Racah matrix $U_{ij}$ must be determined using the above equation. So far,
$U$ up to the size $5\times 5$\cite{Wenzl,IMMMec} and for the size $6\times 6$ \cite{Univ} has been obtained.

Going beyond three-strand braids have not been discussed within the context of eigenvalue hypothesis. In this paper, we demonstrate that the eigenvalue hypothesis is still true for the case of multi-strand braids. We illustrate it in the example of four-strand braid (i.e for the maps $V^{\otimes 4}\to Q$).  In this case there are three ${\cal R}_{i=1,2,3}$-matrices and two unitary matrices
$U,W_1$.
Hereafter, we  call  all these unitary matrices as inclusive Racah matrices even though they are more general matrices that relate the braiding matrices (${\cal R}_i$'s) in the multistrand braids.
Note that the matrix $U$  that make the rotation of ${\cal R}_1$ to ${\cal R}_2$ are still the same and are determined by the Yang-Baxter equation (\ref{YB}), the rotation to $\mathcal{R}_3=UW_1U\mathcal{R}_1U^{\dagger}W_1^{\dagger}U^{\dagger}$ involves also $W_1$ matrix besides $U$, which is determined from the requirement that ${\cal R}_1$ and ${\cal R}_3$ commute:
\begin{equation}
\phantom{.}[{\cal R}_1,{\cal R}_3]=0
\end{equation}
This property comes in fact from the braid-group relations which $\mathcal{R}$-matrices should satisfy. We solved this equation for $W_1$ up to matrices of size $6\times 6$ and checked that the result for the Racah matrices obtained this way coincides with the Racah matrices evaluated by the highest weight method.

With the above discussions for three and four-strand braids, we can now formulate the eigenvalue hypothesis for the generic $n$-strand braid:

\bigskip

\fbox{\parbox{16cm}{{\bf Extended eigenvalue hypothesis.} The Racah matrices defining ${\cal R}_2$ are determined by the Yang-Baxter equation (\ref{YB}), while the remaining Racah matrices defining ${\cal R}_i$, $i\ge 3$ are determined by commuting with all non-neighbour $\mathcal{R}$-matrices (parameterized like ${\cal R}_4$ in (\ref{R4})):
\begin{equation}
\mathcal{R}_i\mathcal{R}_j=\mathcal{R}_j\mathcal{R}_i,\ \ \ |j-i|\neq 1.\nonumber
\end{equation}}}

In other words, according to this hypothesis, if one makes $\mathcal{R}_1$-matrix diagonal with all the eigenvalues different from each other, then all other matrices are uniquely defined. Therefore they provide some particular representation of the braid group.

\bigskip

The paper is organized as follows. In section \ref{s.Rm}, we discuss general properties of $\mathcal{R}$-matrices involved in the Reshetikhin-Turaev approach. In section \ref{s.eig}, we outline the basics of eigenvalue hypothesis with section \ref{s.eig3} presenting old results from \cite{IMMMec} for the 3-strand case and section \ref{s.eig4} presenting new results for 4-strand eigenvalue hypothesis. In section \ref{s.hwm}, we explain the highest weight method, which allows us to evaluate the Racah matrices. Particularly, we indicate the  calculation of $U$ in the three-strand braid for a specific representation. These matrix elements agree with those of the eigenvalue hypothesis in sec.\ref{s.eig} upto $\pm$ sign. We have presented $U$ and $W_1$ for $R=[1]$ in section \ref{s.symrac}. Few of the unitary matrices for $R=[2]$ are given in sec \ref{s.symrac1}. Summary and open questions are posed in the concluding section \ref{conc}. In Appendix \ref{app}, we have placed the remaining $U$ and $W_1$ for $k \times k$ where $k >3$ for representation $R=[2]$. Appendix \ref{polex}  contains $[2]$-colored HOMFLY-PT for few arborescent knots and all non-arborescent knots upto 10-crossings obtained from four-strand braids. We will update these polynomials in our website\cite{knotebook}.

\section{$\mathcal{R}$-matrices \label{s.Rm}}.

One of the most useful approaches to calculate knot polynomials and the one relevant to the subject of the paper is the so-called Reshetikhin-Turaev (RT) approach \cite{RT,MMMII,IMMM3,MMMS21,MMMS31,SSh,RTmod1,RTmod2}. This approach deals with the braid representation of the knot. Each crossing in the braid corresponds to a particular $\mathcal{R}$-matrix. Then the knot polynomial is presented as a character expansion
\begin{equation}
H^K_Y=\sum\limits_{Q\vdash Y^{\otimes m}} S^*_Q C_Q,
\end{equation}
where the sum is over all irreducible representations in the product or representations corresponding to individual components, $m$ is the number of strands, $Y$ is representation on each strand (in this paper, we consider only knots, but most of the formulae in this section can be extended also to links), $C_Q$ is the trace of product of all $\mathcal{R}$-matrices along the braid in the linear space of all intertwining operators $Y^{\otimes m}\to Q$, $S^*_Q$ is the quantum dimension of the representation $Q$.

Let us denote through $\mathcal{R}_i$ the $\mathcal{R}$-matrix corresponding to the crossing between $i$-th and $(i+1)$-th braid. This matrix is defined by the following three properties:
\begin{itemize}
\item The property of any $\mathcal{R}_i$, its characteristic equation\footnote{Since the braid $\mathcal{R}$-matrix acting on $Y\otimes Y$ commutes with the co-product \cite{FRT}, it is diagonal in the basis of irreps $Q\in Y\otimes Y$. Its eigenvalues are all expressed through the basic quantity
$$\varkappa_Q=1/2\sum_i Q_i (Q_i+1-2i)$$
associated with the Young diagram $Q$ with lines $Q_1\ge Q_2\ldots\ge 0$. Hereafter, we don't differ between the representation $Q$ and the Young diagram that describes $Q$. The eigenvalues then are given by the formula
$$\lambda_Q=\epsilon_Q q^{\varkappa_Q}$$
where the sign factors $\epsilon_Q=\pm 1$ depend on whether $Q$ lies in the symmetric ($+1$) or antisymmetric ($-1$) square of $Y$, see \cite{MMMS21,Rama2}.}:
\begin{equation}
\prod\limits_j(\mathcal{R}_i-\lambda_j)=0.
\label{charR}
\end{equation}
\item The Yang-Baxter equation, which, in the case of a braid, has the following form:
\begin{equation}
\mathcal{R}_i\mathcal{R}_{i+1}\mathcal{R}_i=\mathcal{R}_{i+1}\mathcal{R}_{i}\mathcal{R}_{i+1}.
\label{YBi}
\end{equation}
\item The commutativity of non-neighbour $\mathcal{R}$-matrices:
\begin{equation}
\mathcal{R}_i\mathcal{R}_j=\mathcal{R}_j\mathcal{R}_i,\ \ i\neq j\pm 1.
\label{comR}
\end{equation}
\end{itemize}

\subsection{Racah matrices \label{UVW}}

Since all $\mathcal{R}$-matrices in the braid have the same sets of eigenvalues they are related by rotation matrices which are, in fact, the inclusive Racah matrices. These inclusive Racah matrices possess a very special structure.

Let us choose $\mathcal{R}_1$ to be diagonal. $\mathcal{R}_1$ can be associated with the following way of putting parentheses in the product of representations: $(...((Y\otimes Y)\otimes Y)\otimes..Y)$. Then, $\mathcal{R}_2$ would correspond to the other way: $(...(Y\otimes (Y\otimes Y))\otimes..Y)$. The rotation from one way to another can be described by the following trees of representations:
\begin{equation}
\begin{picture}(200,75)(0,-35)
\put(0,0){\line(-1,1){30}}
\put(0,0){\line(1,1){30}}
\put(-15,15){\line(1,1){15}}
\put(-33,32){\mbox{$Y$}}
\put(-3,32){\mbox{$Y$}}
\put(27,32){\mbox{$Y$}}
\put(-20,0){\mbox{$T$}}
\multiput(3,-3)(3,-3){4}{\circle*{2}}
\put(-5,-15){\mbox{$Q^{\prime}$}}
\put(15,-15){\line(1,1){45}}
\put(57,32){\mbox{$Y$}}
\put(15,-15){\line(1,-1){15}}
\put(27,-40){\mbox{Q}}
\put(55,0){\mbox{$= \ \ \ \sum_{T'}\  u^T_{T^{\prime}} $}}
\put(150,0){
\put(0,0){\line(-1,1){30}}
\put(0,0){\line(1,1){30}}
\put(15,15){\line(-1,1){15}}
\put(-33,32){\mbox{$Y$}}
\put(-3,32){\mbox{$Y$}}
\put(27,32){\mbox{$Y$}}
\put(10,0){\mbox{$T^{\prime}$}}
\multiput(3,-3)(3,-3){4}{\circle*{2}}
\put(-5,-15){\mbox{$Q^{\prime}$}}
\put(15,-15){\line(1,1){45}}
\put(57,32){\mbox{$Y$}}
\put(15,-15){\line(1,-1){15}}
\put(27,-40){\mbox{Q}}
}
\end{picture}
\end{equation}
$u^T_{T^{\prime}}$ are elements of the matrix $U$ corresponding to the Racah coefficient $\left[ \begin{array}{ccc} Y&Y&T \\ Y & Q^{\prime} & T^{\prime} \end{array} \right]$. $\mathcal{R}_2$ is then defined as

\begin{equation}
\mathcal{R}_2=U\mathcal{R}_1U^{\dagger}.
\end{equation}

If one studies three-strand braids then $Q^{\prime}=Q$. However, for larger number of strands, they are not equal. Thus, from the form of this inclusive Racah matrix it is obvious that only the elements of the matrix corresponding to the same $Q^{\prime}$ are non-zero. Hence, such $U$ has a block diagonal form with different blocks corresponding to different $Q^{\prime}$.

The third matrix, $\mathcal{R}_3$ corresponds to the following product of representations: $(...(Y\otimes(Y\otimes (Y\otimes Y)))\otimes..Y)$. Thus, the transition from $\mathcal{R}_1$ to $\mathcal{R}_3$ should be made through the chain of trees

\begin{equation}
\begin{picture}(400,70)(-40,-30)
\put(0,0){\line(-1,1){30}}
\put(0,0){\line(1,1){30}}
\put(-10,10){\line(1,1){20}}
\put(-20,20){\line(1,1){10}}
\put(-33,32){\mbox{$Y$}}
\put(-13,32){\mbox{$Y$}}
\put(7,32){\mbox{$Y$}}
\put(27,32){\mbox{$Y$}}
\multiput(2,-2)(3,-3){3}{\circle*{2}}
\put(-5,-15){\mbox{$Q^{\prime}$}}
\put(10,-10){\line(1,1){40}}
\put(47,32){\mbox{$Y$}}
\put(10,-10){\line(1,-1){10}}
\put(17,-30){\mbox{$Q$}}
%
\put(45,0){\vector(1,0){25}}
\put(55,-10){\mbox{$U$}}
\put(100,0){
\put(0,0){\line(-1,1){30}}
\put(0,0){\line(1,1){30}}
\put(-10,10){\line(1,1){20}}
\put(0,20){\line(-1,1){10}}
\put(-33,32){\mbox{$Y$}}
\put(-13,32){\mbox{$Y$}}
\put(7,32){\mbox{$Y$}}
\put(27,32){\mbox{$Y$}}
\put(-5,7){\mbox{$T_1$}}
\put(-15,-5){\mbox{$T_2$}}
\multiput(2,-2)(3,-3){3}{\circle*{2}}
\put(-5,-15){\mbox{$Q^{\prime}$}}
\put(10,-10){\line(1,1){40}}
\put(47,32){\mbox{$Y$}}
\put(10,-10){\line(1,-1){10}}
\put(17,-30){\mbox{$Q$}}
}
\put(145,0){\vector(1,0){25}}
\put(155,-10){\mbox{$W_{1}$}}
\put(200,0){
\put(0,0){\line(-1,1){30}}
\put(0,0){\line(1,1){30}}
\put(10,10){\line(-1,1){20}}
\put(0,20){\line(1,1){10}}
\put(-33,32){\mbox{$Y$}}
\put(-13,32){\mbox{$Y$}}
\put(7,32){\mbox{$Y$}}
\put(27,32){\mbox{$Y$}}
\put(-5,7){\mbox{$T_1$}}
\put(5,-3){\mbox{$T^{\prime}_2$}}
\multiput(2,-2)(3,-3){3}{\circle*{2}}
\put(-5,-15){\mbox{$Q^{\prime}$}}
\put(10,-10){\line(1,1){40}}
\put(47,32){\mbox{$Y$}}
\put(10,-10){\line(1,-1){10}}
\put(17,-30){\mbox{$Q$}}
}
\put(245,0){\vector(1,0){25}}
\put(255,-10){\mbox{$U$}}
\put(300,0){
\put(0,0){\line(-1,1){30}}
\put(0,0){\line(1,1){30}}
\put(10,10){\line(-1,1){20}}
\put(20,20){\line(-1,1){10}}
\put(-33,32){\mbox{$Y$}}
\put(-13,32){\mbox{$Y$}}
\put(7,32){\mbox{$Y$}}
\put(27,32){\mbox{$Y$}}
\multiput(2,-2)(3,-3){3}{\circle*{2}}
\put(-5,-15){\mbox{$Q^{\prime}$}}
\put(10,-10){\line(1,1){40}}
\put(47,32){\mbox{$Y$}}
\put(10,-10){\line(1,-1){10}}
\put(17,-30){\mbox{$Q$}}
}
\end{picture}
\label{eqtrees}
\end{equation}

The first and the last matrices are actually the same matrices as appeared above. But for relating $\mathcal{R}_3$  with $\mathcal{R}_1$ a new matrix $W_1$ is needed. This matrix corresponds to the Racah coefficient $\left[ \begin{array}{ccc} Y&T_1&T_2 \\ Y & Q^{\prime} & T_2^{\prime} \end{array} \right]$.  $\mathcal{R}_3$ is then defined as

\begin{equation}
\mathcal{R}_3=UW_1U\mathcal{R}_1U^{\dagger}W_1^{\dagger}U^{\dagger}.
\end{equation}

Again, similarly to the case of $U$, the only non-zero elements of matrix $W_1$ correspond to the coinciding $T_1$ and coinciding $Q^{\prime}$. This leads to a very interesting property of the matrix $W_1$. Since the eigenvalues of the diagonal $\mathcal{R}$-matrix are defined by the representation $T_1$, $W_1$ commutes with the diagonal $\mathcal{R}$-matrix.

Similarly, this structure can be continued to further $\mathcal{R}$-matrices, e.g. for $\mathcal{R}_4$:
\begin{equation}\label{R4}
\mathcal{R}_4=UW_1UW_2UW_1U\mathcal{R}_1U^{\dagger}W_1^{\dagger}U^{\dagger}W_2^{\dagger}U^{\dagger}W_1^{\dagger}U^{\dagger}.
\end{equation}
$W_2$ then again possesses a block structure. This block structure can be described by paths from the initial representation $Y$ to the final representation $Q$ \cite{MMMII}. This is a generalization of the statements made about the block structure of matrices $U$ and $W_1$. If one defines the representation $Q$ as coming from the following sequence of representations
\begin{equation}
Y\rightarrow T_1 \rightarrow T_2 \rightarrow T_3 \rightarrow ... \rightarrow Q,
\end{equation}
then the matrix $W_i$ has non-zero elements only for the final representations $Q$ corresponding to the same $T_1$, $T_2$,.. $T_i$, $T_{i+2}$, ... This allows one to define the block structure of any Racah matrix $W_i$.

Let us discuss some particular example of this path and block structure, e.g. $[2]^{\otimes 4}=[5,3]$. This is a $4$-strand case, thus, only the Racah matrices $U$ and $W_1$ appear. The multiplicity of representation $[5,3]$ is equal to $6$ which means $6$ possible paths:
\begin{equation}
\begin{array}{llclclcl}
1. & [2] & \rightarrow & [4] & \rightarrow & [5,1] & \rightarrow & [5,3] \\
2. & [2] & \rightarrow & [4] & \rightarrow & [4,2] & \rightarrow & [5,3] \\
3. & [2] & \rightarrow & [3,1] & \rightarrow & [5,1] & \rightarrow & [5,3] \\
4. & [2] & \rightarrow & [3,1] & \rightarrow & [4,2] & \rightarrow & [5,3] \\
5. & [2] & \rightarrow & [3,1] & \rightarrow & [3,3] & \rightarrow & [5,3] \\
6. & [2] & \rightarrow & [2,2] & \rightarrow & [4,2] & \rightarrow & [5,3]
\end{array}
\end{equation}
The matrix $U$ then has three blocks. The first one mixes paths $1$ and $3$, the second one mixes paths $2$, $4$ and $6$ and the third one corresponds only to path $5$. Then the matrix $W_1$ also has three blocks. The first one mixes paths $1$ and $2$, the second one mixes paths $3$, $4$ and $5$ and the third one corresponds to path $6$. 

\section{Eigenvalue hypothesis \label{s.eig}}

In this section, we discuss how the properties of $\mathcal{R}$-matrices define their form and how the eigenvalue hypothesis appear from these properties.

\subsection{2-strand case}

In the 2-strand case, there exists only one $\mathcal{R}$-matrix and only one property of the three discussed is important, namely, the characteristic equation (\ref{charR}). This property defines eigenvalues of the $\mathcal{R}$-matrix. In fact, this property is essentially 2-strand. This means that even if we study larger number of strands it does not give any further information and includes only one $\mathcal{R}$-matrix acting on two adjacent strands.

\subsection{3-strand case \label{s.eig3}}

In the 3-strand case, there are two $\mathcal{R}$-matrices: $\mathcal{R}_1$ and $\mathcal{R}_2$, related by one Racah matrix $U$:
\begin{equation}
\mathcal{R}_2=U\mathcal{R}_1 U^{\dagger}
\end{equation}
As already explained in the previous subsection, (\ref{charR}) does not give any new information about these $\mathcal{R}$-matrices and only describes that they have the same eigenvalues. However, the Yang-Baxter equation (\ref{YBi}) is of great importance here. In terms of the Racah matrix $U$ this equation looks like
\begin{equation}
\mathcal{R}_1 U \mathcal{R}_1 U^{\dagger} \mathcal{R}_1=U \mathcal{R}_1 U^{\dagger} \mathcal{R}_1 U \mathcal{R}_1 U^{\dagger}
\label{YB15}
\end{equation}
Incorporating unitarity of  $U$ ( $UU^{\dagger}=1$) into the above equation one comes to the eigenvalue hypothesis. For the matrices of the size up to $6\times 6$, there is a \textbf{unique} solution for matrix $U$ if $\mathcal{R}$-matrix is chosen to be diagonal. Strictly speaking, there are several solutions differing by inessential sign changes, and, at some special values of the eigenvalues, more solutions can also emerge. For $2\times 2$ matrices the unique solution to (\ref{YB15}) in the case of two generic eigenvalues looks like:
\begin{equation}
\label{Um22}
U=\left(\begin{array}{cc}
\cfrac{\sqrt{-\lambda_1\lambda_2}}{\lambda_1-\lambda_2} & \cfrac{\sqrt{\lambda_1^2-\lambda_1\lambda_2+\lambda_2^2}}{\lambda_1-\lambda_2}
\\
\cfrac{\sqrt{\lambda_1^2-\lambda_1\lambda_2+\lambda_2^2}}{\lambda_1-\lambda_2} & \cfrac{\sqrt{-\lambda_1\lambda_2}}{\lambda_1-\lambda_2}
\end{array}\right),\ \text{for}\
\mathcal{R}_1=\left(\begin{array}{cc}
\lambda_1 &
\\
 & \lambda_2
\end{array}\right).
\end{equation}

Similar answers can be found for matrices of larger sizes.

As characteristic equation was essentially a 2-strand property, the Yang-Baxter equation is a 3-strand property and does not give anything new for larger number of strands. This is explained in detail for the 4-strand braid in the next subsection.

\subsection{4-strand case\label{s.eig4}}

For $4$-strand situation there are three $\mathcal{R}$ matrices and two inclusive Racah matrices, $U$ and $W_1$. Here, the most important is the third property (\ref{comR}).

Suppose all the eigenvalues of $\mathcal{R}$-matrices are different, then the only solution to (\ref{comR}) is to have $\mathcal{R}_1=\mathcal{R}_3$ and $\mathcal{R}_2$ is defined from first three strands as discussed in subsection \ref{s.eig3}.

If some eigenvalues coincide the situation is more interesting. According to subsection \ref{UVW}, both $U$ and $W_1$ have the block-diagonal form. Also $W_1$ commutes with $\mathcal{R}_1$. If $\mathcal{R}_1$ is diagonal, then
\begin{equation}
\mathcal{R}_2=U\mathcal{R}_1U^{\dagger},\ \ \ \mathcal{R}_3=UW_1U\mathcal{R}_1U^{\dagger}W_1^{\dagger}U^{\dagger},
\end{equation}
on the other hand, if one diagonalizes $\mathcal{R}_2=\mathcal{R}$, then $\mathcal{R}_3=W_1U\mathcal{R}_2U^{\dagger}W_1^{\dagger}$, and $W_1$ commutes with $\mathcal{R}$. Then, the Yang-Baxter equation on $\mathcal{R}_2$ and $\mathcal{R}_3$ looks like
\begin{equation}
\mathcal{R}\left(W_1U\mathcal{R}U^{\dagger}W_1^{\dagger}\right)\mathcal{R}=\left(W_1U\mathcal{R}U^{\dagger}W_1^{\dagger}\right)\mathcal{R}\left(W_1U\mathcal{R}U^{\dagger}W_1^{\dagger}\right).
\end{equation}
Since $W_1$ commutes with $\mathcal{R}$, this equation transforms into
\begin{equation}
\mathcal{R}U \mathcal{R} U^{\dagger} \mathcal{R}=U \mathcal{R} U^{\dagger} \mathcal{R} U \mathcal{R} U^{\dagger},
\end{equation}
which is automatically satisfied because of the construction in section \ref{s.eig3} and does not include $W_1$. Thus, $W_1$ cannot be found from the Yang-Baxter equation, and one needs another equation for the Racah coefficients to find $W_1$. This equation comes from the third property (\ref{comR}):
\begin{equation}
\label{Weq}
UW_1U\mathcal{R}U^{\dagger}W_1^{\dagger}U^{\dagger}\mathcal{R}=\mathcal{R}UW_1U\mathcal{R}U^{\dagger}W_1^{\dagger}U^{\dagger}
\end{equation}
\begin{itemize}
\item[1)]
If the matrix is of size $2\times 2 $, or if the matrix $U$ mixes all the eigenvalues, then the only solution for the matrix $W_1$ is identity matrix.
\item[2)]
If the matrix is of size $3\times 3$, and the diagonal matrix $\mathcal{R}$ looks like
\begin{equation}
\mathcal{R}_1=\left(\begin{array}{ccc}
\lambda_1 &&
\\
& \lambda_1 &
\\
&& \lambda_2
\end{array}\right)
\end{equation}
then
\begin{equation}
U=\left(\begin{array}{ccc}
1&& \\
& \cfrac{\sqrt{-\lambda_1\lambda_2}}{\lambda_1-\lambda_2} & \cfrac{\sqrt{\lambda_1^2-\lambda_1\lambda_2+\lambda_2^2}}{\lambda_1-\lambda_2}
\\
& \cfrac{\sqrt{\lambda_1^2-\lambda_1\lambda_2+\lambda_2^2}}{\lambda_1-\lambda_2} & -\cfrac{\sqrt{-\lambda_1\lambda_2}}{\lambda_1-\lambda_2}
\end{array}\right)
\label{Um33}
\end{equation}
and
\begin{equation}
W_1=\left(\begin{array}{ccc}
-\cfrac{\lambda_1\lambda_2}{\lambda_1^2-\lambda_1\lambda_2+\lambda_2^2} & \cfrac{\sqrt{\lambda_1^2+\lambda_2^2}(\lambda_1-\lambda_2)}{\lambda_1^2-\lambda_1\lambda_2+\lambda_2^2} &
\\
\cfrac{\sqrt{\lambda_1^2+\lambda_2^2}(\lambda_1-\lambda_2)}{\lambda_1^2-\lambda_1\lambda_2+\lambda_2^2} & \cfrac{\lambda_1\lambda_2}{\lambda_1^2-\lambda_1\lambda_2+\lambda_2^2} &
\\ && 1
\end{array}\right)
\label{Wm33}
\end{equation}
If the diagonal $\mathcal{R}$-matrix has a different order of eigenvalues or the paths go differently mixing the first and the last eigenvalues, this leads to permutations of rows and columns in the $U$ and $W_1$ matrices, but the same formulae still work.

\item[3)]
For the matrix of the size $6\times 6$ appearing in the product of four representations $[2]$, the situation is like this. The initial diagonal matrix is of the form $\mathcal{R}=diag(\lambda_1,\lambda_1,\lambda_2,\lambda_1,\lambda_2,\lambda_3)$ and the $U$-matrix consists of three blocks of sizes $1\times 1$, $2\times 2$ and $3\times 3$. Then the matrix elements from the eigenvalue hypothesis are:
\setlength{\arraycolsep}{1pt}
\begin{equation}
\!\!\!\!\!\!\!\!\!\!\!\!\!\!\!
U=\left(
\begin{array}{cccccc}
1 \\
& \frac{\sqrt{-\lambda_1\lambda_2}}{\lambda_1-\lambda_2} & \frac{\sqrt{\lambda_1^2-\lambda_1\lambda_2+\lambda_2^2}}{\lambda_1-\lambda_2} \\
& \frac{\sqrt{\lambda_1^2-\lambda_1\lambda_2+\lambda_2^2}}{\lambda_1-\lambda_2} & -\frac{\sqrt{-\lambda_1\lambda_2}}{\lambda_1-\lambda_2} \\
&&& \frac{\lambda_1(\lambda_2+\lambda_3)}{(\lambda_1-\lambda_3)(\lambda_1-\lambda_2)} &
\frac{\sqrt{(\lambda_1^2+\lambda_2\lambda_3)(\lambda_2^2+\lambda_1\lambda_3)}}{(\lambda_1-\lambda_2)\sqrt{(\lambda_1-\lambda_3)(\lambda_3-\lambda_2)}} &
\frac{\sqrt{(\lambda_1^2+\lambda_2\lambda_3)(\lambda_3^2+\lambda_1\lambda_2)}}{(\lambda_1-\lambda_3)\sqrt{(\lambda_1-\lambda_2)(\lambda_2-\lambda_3)}} \\
&&& \frac{\sqrt{(\lambda_1^2+\lambda_2\lambda_3)(\lambda_2^2+\lambda_1\lambda_3)}}{(\lambda_1-\lambda_2)\sqrt{(\lambda_1-\lambda_3)(\lambda_3-\lambda_2)}} &
-\frac{\lambda_2(\lambda_1+\lambda_3)}{(\lambda_1-\lambda_2)(\lambda_2-\lambda_3)} &
\frac{\sqrt{(\lambda_2^2+\lambda_1\lambda_3)(\lambda_3^2+\lambda_1\lambda_2)}}{(\lambda_2-\lambda_3)\sqrt{-(\lambda_1-\lambda_2)(\lambda_1-\lambda_3)}} \\
&&& \frac{\sqrt{(\lambda_1^2+\lambda_2\lambda_3)(\lambda_3^2+\lambda_1\lambda_2)}}{(\lambda_1-\lambda_3)\sqrt{(\lambda_1-\lambda_2)(\lambda_2-\lambda_3)}} &
\frac{\sqrt{(\lambda_2^2+\lambda_1\lambda_3)(\lambda_3^2+\lambda_1\lambda_2)}}{(\lambda_2-\lambda_3)\sqrt{-(\lambda_1-\lambda_2)(\lambda_1-\lambda_3)}} &
\frac{\lambda_3(\lambda_1+\lambda_2)}{(\lambda_1-\lambda_3)(\lambda_2-\lambda_3)}
\end{array}
\right)
\label{U66}
\end{equation}
\setlength{\arraycolsep}{6pt}
Then from (\ref{Weq}), one can find how the matrix $W_1$ looks like:
\begin{equation}
W_1=\left(
\begin{array}{cccccc}
A_6 & D_6 &  & E_6 \\
D_6 & B_6 &  & F_6 \\
    &     & \frac{\lambda_3(\lambda_1+\lambda_2)}{\lambda_1\lambda_2+\lambda_3^2} & & \frac{\sqrt{(\lambda_1^2-\lambda_3^2)(\lambda_2^2-\lambda_3^2)}}{\lambda_3^2+\lambda_1\lambda_2} \\
E_6 & F_6 &  & C_6 \\
    &     & \frac{\sqrt{(\lambda_1^2-\lambda_3^2)(\lambda_2^2-\lambda_3^2)}}{\lambda_3^2+\lambda_1\lambda_2} & & -\frac{\lambda_3(\lambda_1+\lambda_2)}{\lambda_1\lambda_2+\lambda_3^2} \\
    &     &  &                                                                    & &  1
\end{array}
\right),
\label{W66}
\end{equation}
where
\begin{equation}
\begin{array}{ll}
A_6=\frac{\lambda_1^2\lambda_2(\lambda_2+\lambda_3)}{(\lambda_1^2+\lambda_2\lambda_3)(\lambda_1^2-\lambda_1\lambda_2+\lambda_2^2)};
&
B_6=\frac{\lambda_2(\lambda_1^3-\lambda_1^2\lambda_2-\lambda_1^2\lambda_3-\lambda_1\lambda_3^2+\lambda_2\lambda_3^2-\lambda_1\lambda_2\lambda_3)}{(\lambda_3^2+\lambda_1\lambda_2)(\lambda_1^2-\lambda_1\lambda_2+\lambda_2^2)};
\\
C_6=\frac{\lambda_1\lambda_3(\lambda_1+\lambda_2)(\lambda_2+\lambda_3)}{(\lambda_1^2+\lambda_2\lambda_3)(\lambda_3^2+\lambda_1\lambda_2)};
&
D_6=\frac{1}{\lambda_1^2-\lambda_1\lambda_2+\lambda_2^2}\sqrt\frac{\lambda_1(\lambda_1-\lambda_2)(\lambda_2+\lambda_3)(\lambda_2^2+\lambda_1\lambda_3)(\lambda_1^3-\lambda_2^2\lambda_3)}{(\lambda_1^2+\lambda_2\lambda_3)(\lambda_3^2+\lambda_1\lambda_2)};
\\
E_6=-\frac{\lambda_1-\lambda_3}{\lambda_1^2+\lambda_2\lambda_3}\sqrt{\frac{\lambda_2(\lambda_1+\lambda_3)(\lambda_1-\lambda_2)(\lambda_1^3-\lambda_2^2\lambda_3)}{(\lambda_3^2+\lambda_1\lambda_2)(\lambda_1^2-\lambda_1\lambda_2+\lambda_2^2)}};
&
F_6=-\frac{\lambda_1-\lambda_3}{\lambda_3^2+\lambda_1\lambda_2}\sqrt{\frac{\lambda_1\lambda_2(\lambda_1+\lambda_3)(\lambda_2+\lambda_3)(\lambda_2^2+\lambda_1\lambda_3)}{(\lambda_1^2+\lambda_2\lambda_3)(\lambda_1^2-\lambda_1\lambda_2+\lambda_2^2)}}.
\end{array}
\end{equation}
\item[4)]
For larger sizes of matrices, the blocks of size $4\times 4$ appear, and it is rather tedious to find the answers in this case.
\end{itemize}
The results of this section have been confirmed from the highest weight method (reviewed in the next section \ref{s.hwm}) for some representations which are presented in section \ref{s.symrac} and Appendix \ref{app}.



\section{Highest weight method\label{s.hwm}}
In this section, we formally present the highest weight method for any $m$-strand braid carrying an arbitrary representation $Y$ of quantum group $SU_q(N)$.

This method is a systematic procedure which allows one to construct a highest weight vector state. It is based on the manifest action of lowering $T_{i}^{-}$and raising operators $T_{i}^{+}$ on representations of $SU_{q}(N)$ \cite{MMMII}:
\begin{equation}
\begin{array}{lcr}
T^{-}_{i}V_i =V_{i-1};&\ & T^{+}_{i}V_{i-1}=V_{i}.\\
H_{i}V_{i}=+\frac{1}{2}V_{i};&\ &
H_{i}V_{i-1}=-\frac{1}{2}V_{i-1}.
\end{array}
\end{equation}
where $V_i$ is an $i$-th vector of the fundamental representation, and $T_{i}^{+}$, $T_{i}^{-}$ and $q^{H_{i}}$ are generators of $SU_{q}(N)$. To generalize this action to higher rank tensors, one has to define a comultiplication $\Delta$ on $SU_{q}(N)$:
\begin{equation}
\begin{array}{lcl}
\Delta(T^{+}_{i})&=& \mathbb{I}  \otimes  T^{+}_{i}  + T^{+}_{i}\otimes q^{-2 H_{i}}  \\
\Delta(T^{-}_{i})&=&q^{2 H_{i}}  \otimes T^{-}_{i}+ T^{-}_{i}\otimes     \mathbb{I}.
\end{array}
\end{equation}
This extends the action of $T^{\pm}_{i}$ to tensors of any rank. We indicate the highest weight vector for a representation $R$ labeled by Young diagram $[\lambda_1,\lambda_2,\lambda_3 \ldots \lambda_l]$ as a  sum of  $V_{(\underbrace{0,0,0}_{\lambda_1},\ldots,\underbrace{1,1,1}_{\lambda_2},\ldots)}$ and their permutations.  \\
For example, the highest weight state for $R=[1,1]$ will involve $V_{0,1}$ and $V_{1,0}$. One can construct all the highest weight vectors of the representation by using the lowering and raising operators $T^{\pm}_{i}$ and the comultiplication rule.


We would like to validate the results of inclusive Racah matrices obtained from eigenvalue hypothesis  using the highest weight approach. For definiteness, we take representation $R=[2]$ and construct the highest weight states for $m=3$ and $m=4$ strand braids in the following subsections.



\subsection{Evaluation of vector states for $m=3$}

Our goal is to evaluate the $U$ matrix corresponding to all irreducible representations in the fusion channel of $[2]^{\otimes 3}$ for $m=3$ strand braid shown in eqn.(29). One can easily see the two possible fusion trees which give two sets of irreducible representations $Q$ in the final channel which are related by a unitary matrix, its size being determined by the multiplicity indicated by red color.
\begin{equation}
\begin{picture}(300,70)(-10,-10)
\put(0,0){\line(-1,1){30}}
\put(-10,10){\line(1,1){20}}
\put(-20,20){\line(1,1){10}}
\put(-33,33){\mbox{$[2]$}}
\put(-13,33){\mbox{$[2]$}}
\put(7,33){\mbox{$[2]$}}
\put(-24,6){\mbox{$T$}}
\put(-12,-7){\mbox{$Q$}}

\put(-32,-24){\mbox{$\{ ([2] \otimes[2])_{T}\otimes [2] \}^{R}_{Q}$}}
%
\put(90,10){\vector(1,0){50}}
\put(110,-10){\mbox{$U_{Q}$}}
\put(250,0){
\put(0,0){\line(-1,1){30}}
\put(-10,10){\line(1,1){20}}
\put(0,20){\line(-1,1){10}}
\put(-33,33){\mbox{$[2]$}}
\put(-13,33){\mbox{$[2]$}}
\put(7,33){\mbox{$[2]$}}
\put(-3,8){\mbox{$T^{'}$}}
\put(-13,-7){\mbox{$Q$}}
\put(-32,-24){\mbox{$\{ [2] \otimes([2]\otimes [2])_{T'}\}^{L}_{Q}$}}
}
\end{picture}
\label{29}
\end{equation}

\vspace{1cm}
 Here   Q  $\in$ $\{[2,2,2], [6,0], [3,3], [4,1,1],{\textcolor{red}{2}}[5,1], {\textcolor{red}{2}}[3,2,1], {\textcolor{red}{3}}[4,2]\}
$ and $T$ , $T^{\prime} \in \{[4], [2,2],[3,1]\}$.
\paragraph{}
For example, order of matrix $U_{[5,1]}$ is $2 \times 2$ and $U_{ [4,2]}$ is $3 \times 3$.
The highest weight vector $H_T$ for $T\in [2]^{\otimes 2}$ can be determined by applying the raising operator $\Delta T_1^+$ on product
$V_{0,0} \otimes V_{0,0}$.  Clearly, $H_{[4]}= V_{0,0,0,0}$. We will now elaborate the steps involved  for $H_{[3,1]}$ for clarity:

\begin{equation}
\begin{array}{lcl}
H_{[3,1]}&=&\alpha \{ V_{0,0} \otimes( \Delta(T^{+}_{1})V_{0,0})\} +\beta\{( \Delta(T^{+}_{1})V_{0,0}) \otimes V_{0,0} \}\\
&=&\alpha \left\{ V_{0,0,0,1}+q V_{0,0,1,0} \right\} +\beta  \left\{ V_{0,1,0,0}+q V_{1,0,0,0} \right\}.
\end{array}
\end{equation}
To find $\alpha$ and $\beta$, one should impose the highest weight vector condition $\Delta(T^{-}_{1})
H_{[3,1]}=0$, which implies
\begin{equation}
\alpha \left\{(q^{-3}+q^{-1}) V_{0,0,0,0} \right\} +\beta  \left\{ (q+q^{-1}) V_{0,0,0,0}  \right\} =0,
\end{equation}
giving the $\alpha=-q^{+2} ,\quad \beta=1$. Hence, the explicit highest vector state is

\begin{eqnarray}
H_{[3,1]} &=& -q^{2} V_{0,0,0,1}-q^{3} V_{0,0,1,0}  +   V_{0,1,0,0}+q V_{1,0,0,0}.
\end{eqnarray}
By a similar procedure, one can obtain the highest weight vector for $H_{[2,2]}$:
\begin{equation}
H_{[2,2]} =-q V_{0, 0, 1, 1} - q^3 V_{0, 0, 1, 1} + V_{0, 1, 0, 1} +  q V_{0, 1, 1, 0} + q V_{1, 0, 0, 1}+ q^2 V_{1, 0, 1, 0}
 - q^{-1} V_{1, 1, 0, 0} - q V_{1, 1, 0, 0}.\\
\end{equation}

Using the result of highest weight vectors of m=2 strand, we can move to $m=3$ strand braid and do the explicit calculation of the elements of $U_{Q}$ matrices.
 For the sake of definiteness, we focus on one of the representation, i.e [5,1] for $m=3$ strands which has multiplicity of $2$ (see (\ref{29})). The representation $[5,1]$ comes from two sectors, the left sector (L), corresponding to $[2]\otimes [4]$ and $[2]\otimes [3,1]$, and the right sector (R), corresponding to $[4]\otimes [2]$ and $[3,1] \otimes [2]$. On representation $Q$, we place a subcript to keep track of the multiplicity and superscript to denote left or right sector.  
Therefore to find the U matrix for the representation [5,1], one has to solve the following equations:
\begin{equation}
\label{34}
\begin{array}{lcl}
H_{[5,1]_{1}^{R}} \in \underbrace{([4])\otimes[2]}_{\text{Right sector}}&=&
\underbrace{\alpha ([2]\otimes[4])+\beta ([2] \otimes [3,1])}_{\text{Left sector}}\\
H_{[5,1]_{2}^{R}}  \in \underbrace{([3,1])\otimes[2]}_{\text{Right sector}}&=& \underbrace{\gamma ([2]\otimes[31])+\delta ([2] \otimes [3,1])}_{\text{Left sector}}
\end{array}
\end{equation}
Now it remains to present the explicit calculations to determine the unknown parameters $\alpha$, $\beta$, $\gamma$, $\delta$:\\
Right Sector:\\
\begin{equation*}
\begin{array}{lcl}
[5,1]_1^R& \in &[4]\otimes[2]\\
H_{[5,1]^{R}_{1}}&=&\beta_{R} \{ V_{0,0,0,0}\otimes(\Delta T^{+}_{1}V_{0,0})\}+\alpha_{R} \{ (\Delta T^{+}_{1}V_{0,0,0,0})\otimes V_{0,0}\}\\
&=&\alpha_{R} \left\{ V_{0,0,0,1,0,0}+q V_{0,0,1,0,0,0}+q^2V_{0,1,0,0,0,0}+q^3V_{1,0,0,0,0,0} \right\} +\\
&&\beta_{R}  \left\{ V_{0,0,0,0,0,1}+q V_{0,0,0,0,1,0} \right\}
\end{array}
\end{equation*}
To find the value of $\alpha_{R}$ and $\beta_{R}$, we apply the highest weight condition
\begin{equation}
\begin{array}{lcl}
\Delta(T^{-}_{1})H_{[5,1]^{R}_{1}}&=&0\\
&\Rightarrow&\alpha_R \left\{(q^{-3}+q^{-1}+q+q^3) V_{0,0,0,0,0,0} \right\} +\beta_R \left\{ (q^{-3}+q^{-5}) V_{0,0,0,0,0,0}  \right\} =0\\
&\Rightarrow &\alpha_R=-q^{-4}\frac{[2]_{q}}{[4]_{q}} ,   \beta_R=1
\end{array}
\end{equation}
Hence the final form for the highest weight vector is
\begin{equation*}
H_{[5,1]^{R}_{1}}=\frac{1}{\mathcal{N_R}}\{-q^{-4}\frac{[2]_{q}}{[4]_{q}} \left\{ V_{0,0,0,1,0,0}+q V_{0,0,1,0,0,0}+q^2V_{0,1,0,0,0,0}
\\+q^3V_{1,0,0,0,0,0} \right\} +  \left\{ V_{0,0,0,0,0,1}+q V_{0,0,0,0,1,0} \right\} \},
\end{equation*}
where the normalization constant is equal to
\begin{equation}
\mathcal{N_R}=\sqrt{  -q^{-1}[6]_{q}\frac{[2]_{q}}{[4]_{q}}}.
\end{equation}
Left Sector:\\
\begin{equation*}
\begin{array}{lcl}
{[5,1]_1^{L}}&\in &[2]\otimes[4]\\
H_{[5,1]^{L}_{1}}&=&\alpha_{L} \{ V_{0,0}\otimes(\Delta T^{+}_{1}V_{0,0,0,0})\}+\beta_{L} \{ (\Delta T^{+}_{1}V_{0,0})\otimes V_{0,0,0,0}\}\\
&=&\alpha_{L} \left\{ V_{0,0,0,0,0,1}+q V_{0,0,0,0,1,0}+q^2V_{0,0,0,1,0,0}+q^3V_{0,0,1,0,0,0} \right\} +
\beta_{L}  \left\{ V_{0,1,0,0,0,0}+q V_{1,0,0,0,0,0} \right\}
\\ \\
\Delta(T^{-}_{1})H_{[5,1]^{L}_{1}}&=&0\\
&\Rightarrow&\alpha_L \left\{(q^{-5}+q^{-3}+q^{-1}+q) V_{0,0,0,0,0,0} \right\} +\beta_L  \left\{ (q^{-1}+q) V_{0,0,0,0,0,0}  \right\} =0\\
&\Rightarrow& \alpha_L=-q^{2}\frac{[2]_{q}}{[4]_{q}} ,   \beta_L=1
\end{array}
\end{equation*}
Hence the final form for the highest weight vector is
\begin{equation}
\begin{array}{r}
H_{[5,1]^{L}_{1}}=\frac{1}{\mathcal{N_L}}\{-q^{2}\frac{[2]_{q}}{[4]_{q}} \left\{ V_{0,0,0,0,0,1}+q V_{0,0,0,0,1,0}+q^2V_{0,0,0,1,0,0}+q^3V_{0,0,1,0,0,0} \right\} +\\
\left\{ V_{0,1,0,0,0,0}+q V_{1,0,0,0,0,0} \right\},
\end{array}
\end{equation}
where the normalization constant is equal to
\begin{equation}
\mathcal{N_L}=\sqrt{  -q^{-3}[6]_{q}\frac{[2]_{q}}{[4]_{q}}}.
\end{equation}

Similarly, one can construct the highest weight vectors $H_{[5,1]^{R}_{2}}$ and $H_{[5,1]^{L}_{2}}$ which comes from representation $[3,1]$.
 Hence,
 $\alpha$, $\beta$, $\gamma$ and $\delta$ for the matrix $U_{[5,1]}$ are equal to:

\begin{center}
\begin{tabular}{ |p{2cm}||p{5cm}|p{3cm}|  }
\hline
\hspace{5cm} $ U_{[5,1]} $& $H_{[5,1]^{L}_{1}}$&$H_{[5,1]^{L}_{2}}$ \\
\hline
\hline
 $H_{[5,1]^{R}_{1}}$&$\alpha =\frac{q^2}{(1+q^4)}$ &$\beta =\sqrt{\frac{1+q^4+q^8}{(1+q^4)^2}}$  \\
\hline
$H_{[5,1]^{R}_{2}}$ & $\gamma =-\sqrt{\frac{1+q^4+q^8}{(1+q^4)^2}}$& $\delta =\frac{q^2}{(1+q^4)}$  \\
\hline
\end{tabular}
\end{center}
Similarly one can obtain the $U$ matrices for all the representations in $m=3$ case. Interestingly, the magnitude of all the elements coincide with the result obtained from the eigenvalue hypothesis in equations (\ref{Um22}) and (\ref{Um33}). In the following section, we aim to compare the inclusive Racah matrices with the matrices given in section \ref{s.eig} for $m=4$ strands.

\subsection{Evaluation of $U$ and $W_1$ matrix for $m=4$}

\begin{figure}[h!]
\definecolor{ffqqtt}{rgb}{1.,0.,0.2}
\definecolor{ffqqqq}{rgb}{1.,0.,0.}
\definecolor{sqsqsq}{rgb}{0.12549019607843137,0.12549019607843137,0.12549019607843137}
\definecolor{qqzzff}{rgb}{0.,0.6,1.}
\begin{tikzpicture}[line cap=round,line join=round,>=triangle 45,x=.6
0cm,y=1.0cm,scale=.3]
\clip(8.,-6.) rectangle (120.,23.);
\draw (9.335336945031727,14.0
49996666857155) node[anchor=north west] {$\mathbf{[4,0]}$};
\draw (23.551228417899944,14.496702093213744) node[anchor=north west] {$\mathbf{[3,1]}$};
\draw (38.50723616983793,14.496702093213744) node[anchor=north west] {$\mathbf{[2,2]}$};
\draw [color=qqzzff](23.35180981324878,20.072864883911418) node[anchor=north west] {\textbf{[2]$\otimes$[2]}};
\draw [->,line width=1.pt,color=sqsqsq] (25.37790697674419,18.147653467420902) -- (10.562015503875973,14.150560444165093);
\draw [->,line width=1.pt,color=sqsqsq] (25.37790697674419,18.147653467420902) -- (25.377906976744196,14.257149591451917);
\draw (39.12700361169839,8.054647829647852) node[anchor=north west] {$\mathbf{[4,2]}$};
\draw (24.684174154334052,8.061236976934673) node[anchor=north west] {$\mathbf{[5,1]}$};
\draw (9.068864076814673,8.4600771860655603) node[anchor=north west] {$\mathbf{[6,0]}$};
\draw [color=ffqqqq](84.0552981853418,11.828291240500565) node[anchor=north west] {$\mathbf{[2]^{\otimes3}}$};
\draw (64.03010438689219,11.787012170733123) node[anchor=north west] {$\mathbf{\text{m=3 strand}}$};
\draw [color=ffqqqq](22.938050123326303,8.08640077709867) node[anchor=north west] {\textbf{2}};
\draw [color=ffqqqq](37.27429043340382,8.042050930423045) node[anchor=north west] {\textbf{3}};
\draw (23.36440671247359,-0.08739976763234704) node[anchor=north west] {$\mathbf{[6,2]}$};
\draw [color=ffqqqq](84.02332144115576,5.38739976763234704) node[anchor=north west] {$\mathbf{[2]^{\otimes 4}}$};
\draw [->,line width=1.pt,color=sqsqsq] (25.37790697674419,18.09435889377752) -- (39.980620155038764,14.363738738738757);
\draw [color=ffqqtt](21.40510438689219,-0.0019025579490484) node[anchor=north west] {\textbf{6}};
\draw [->,line width=1.pt,color=sqsqsq] (25.377906976744185,12.76490152943643) -- (25.324612403100776,8.128273622459684);
\draw [->,line width=1.pt,color=sqsqsq] (11.041666666666666,12.445134087575966) -- (25.324612403100776,8.128273622459684);
\draw [->,line width=1.pt,color=sqsqsq] (40.03391472868217,12.65831238214961) -- (25.324612403100776,8.128273622459684);
\draw [->,line width=1.pt,color=sqsqsq] (10.455426356589147,12.338544940289143) -- (10.508720930232558,8.607924785250383);
\draw [->,line width=1.pt,color=sqsqsq] (10.7218992248062,7.222265870521699) -- (25.218023255813954,-0.2922690131992356);
\draw [->,line width=1.pt,color=sqsqsq] (11.319909418973037,12.36103833034904) -- (40.30038759689923,7.968389901529451);
\draw [->,line width=1.pt,color=sqsqsq] (26.017441860465116,12.71160695579302) -- (40.30038759689923,7.968389901529451);
\draw [->,line width=1.pt,color=sqsqsq] (40.35368217054264,12.65831238214961) -- (40.30038759689923,7.968389901529451);
\draw [->,line width=1.pt,color=sqsqsq] (25.271317829457363,6.582730986800769) -- (25.218023255813954,-0.2922690131992356);
\draw [->,line width=1.pt,color=sqsqsq] (39.71414728682171,6.529436413157358) -- (25.218023255813954,-0.2922690131992356);
\draw (64.10471678999296,5.2191827133687814) node[anchor=north west] {$\mathbf{\text{m=4 strand}}$};
\draw (63.18998810782242,16.44883387615948) node[anchor=north west] {$\mathbf{\text{m=2 strand}}$};
\draw [color=ffqqqq](84.0552981853418,16.690112945926922) node[anchor=north west] {$\mathbf{ [2]^{\otimes 2 }}$};
\draw (53.25006562720227,9.787012170733123) node[anchor=north west] {$U_{Q}: \lbrace [2] \otimes([2]\otimes [2])_{\tiny{T}}\rbrace^{\tiny{L}}_{\tiny{Q}}\rightarrow \lbrace ([2] \otimes[2])_{\tiny{T'}}\otimes [2]\rbrace^{\tiny{R}}_{\small Q}$};
\draw (36.71033694503172,2.5922447288726573) node[anchor=north west] {$(UW_{1}U)_{Q}:\{[2] \otimes( [2] \otimes([2]\otimes [2])_{T})^{L}_{Q'})\}^{L}_{Q}\rightarrow \{ (([2] \otimes[2])_{T'}\otimes [2])^{R}_{Q'}\otimes [2])\}^{R}_{Q}$};
\draw (36.29657725510924,-0.4988405424451724) node[anchor=north west] {$(UW_{1})_{Q}:\{[2] \otimes(([2] \otimes[2])_{T'}\otimes [2])^{R}_{Q'}\}^{L}_{Q}\rightarrow \{ (([2] \otimes[2])_{T'}\otimes [2])^{R}_{Q'}\otimes [2])\}^{R}_{Q}$};
\end{tikzpicture}
\caption{Steps for $UW_1U$ unitary matrices computation for $m=4$ strands.}
\label{f.uwu}
\end{figure}
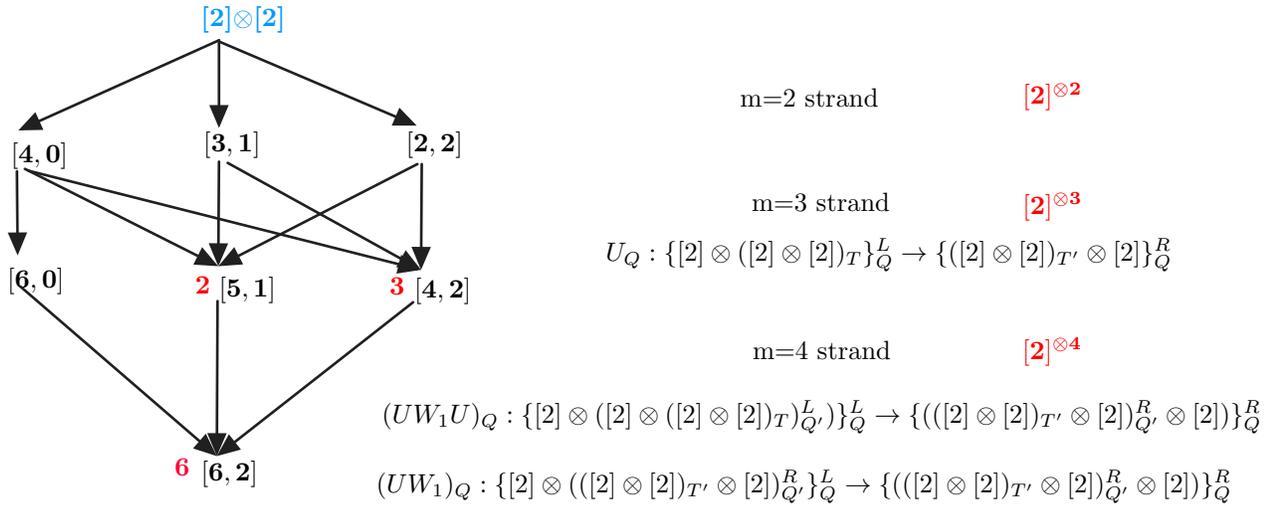

The detailed procedure which we discussed  for determining $U$ matrices  for 3-strand braids can be extended to 4-strand braids. Here we need to consider $Q \vdash [2]^{\otimes 4}$.   The possible representations $Q$ with their multiplicities are tabulated below:

{\small
\begin{center}
\begin{tabular}{ |p{7cm}|p{2cm}|p{2cm}|  }
\hline
\multicolumn{3}{|c|}{m=4 strand braid } \\
\hline
\hspace{3cm}Q    & Matrix size & $\# $ of matrices \\
\hline
[2,2,2,2], [5,1,1,1], [8,0]& 1 &3 \\
\hline
[3,3,1,1] & 2  & 1 \\
\hline
[3,2,2,1], [3,3,2], [4,2,1,1], [4,4], [6,1,1], [7,1]& 3 & 6 \\
\hline
[4,2,2], [5,3], [6,2]  &6 & 3 \\
\hline
[4,3,1] & 7 & 1 \\
\hline
[5,2,1] & 8 & 1   \\

\hline
\end{tabular}
\end{center}
}

From (\ref{eqtrees}), we observe that the matrix $UW_1 U$ relates right sector highest weight vector with the corresponding left sector highest weight vector. We present the calculations for
representation $[6,2]$ (see Fig.\ref{f.uwu}). Let us emphasise that the subsector states involving $3$-strands belong to different sectors. In order to obtain $U$ and $W_1$, we need to  determine the matrix $UW_1$ as well.  In fact, the matrix $UW_1$ relates the two highest weight vectors of different sectors but both their subsectors are either left or right sector. We have highlighted them for representation $[6,2]$ in Fig.\ref{f.uwu}.

It is clear from Fig.\ref{f.uwu} that there are $6$ independent paths to obtain representation $[6,2]$. Therefore, both $UW_1 U$ and $UW_1$ matrices will be of the size $6\times 6$. Following the highest weight method procedure, one can work out the six right sector highest weight vectors $[6,2]^R_i$ and similarly the six left sector highest weight vectors $[6,2]^L_j$, whose subsectors are different. Hence, one can determine the $36$ matrix elements of $UW_1U$ by taking inner product of the left and right sector states.

One could again work out six highest weight states for left and right sectors but with both subsectors being the same. The inner product of such states will give the elements of the $UW_1$ matrix. We present the $U$ and $W_1$   whose matrix sizes are less than $6\times 6$ in the following section. Other matrices of  the sizes $n \times n$ for $n\geq 6$ are presented in Appendix \ref{app}.

\section{Examples\label{s.symrac}}

For clarity, we will first review and give the matrices for the simplest case when the strands carry fundamental representation $R=[1]$. Then, the case $R=[2]$ will be presented.

\subsection{Representation [1]}

In this case, $[1]^{\otimes 4}=[4]+3[3,1]+2[2,2]+3[2,1,1]+[1,1,1,1]$. Thus, there are two matrices of size $1\times 1$, one matrix of size $2\times 2$ and two matrices of size $3\times 3$. The eigenvalues of $\mathcal{R}$-matrices are in this case $\lambda_1=q$, $\lambda_2=-q^{-1}$. The only non-trivial case here are matrices of size $3\times 3$. The $\mathcal{R}$-matrices are equal to:
\begin{equation}
\mathcal{R}^{[3,1]}=\left(
\begin{array}{ccc}
q \\ & q \\ & & -q^{-1}
\end{array}
\right),\ \ \
\mathcal{R}^{[2,1,1]}=\left(
\begin{array}{ccc}
q \\ & -q^{-1} \\ & & -q^{-1}
\end{array}
\right).
\end{equation}
Then, the $U$-matrices are equal to
\begin{equation}
U^{[3,1]}=\left(\begin{array}{ccc}
1 \\
& \frac{1}{[2]} & \frac{\sqrt{[3]}}{[2]} \\
& \frac{\sqrt{[3]}}{[2]} & -\frac{1}{[2]}
\end{array}\right),\ \ \
U^{[2,1,1]}=\left(\begin{array}{ccc}
 \frac{1}{[2]} & \frac{\sqrt{[3]}}{[2]} \\
 \frac{\sqrt{[3]}}{[2]} & -\frac{1}{[2]} \\
 & & 1
\end{array}\right)
\end{equation}
and the $W_1$-matrices are equal to
\begin{equation}
W_1^{[3,1]}=\left(\begin{array}{ccc}
\frac{1}{[3]} & \frac{\sqrt{[2][4]}}{[3]} \\
\frac{\sqrt{[2][4]}}{[3]} & -\frac{1}{[3]} \\
& & 1
\end{array}\right),\ \ \
W_1^{[2,1,1]}=\left(\begin{array}{ccc}
1 \\
& \frac{1}{[3]} & \frac{\sqrt{[2][4]}}{[3]} \\
& \frac{\sqrt{[2][4]}}{[3]} & -\frac{1}{[3]}
\end{array}\right),
\end{equation}
which is in a full accordance with formulae (\ref{Um33}) and (\ref{Wm33}) from section \ref{s.eig3}.

\subsection{Representation [2]\label{s.symrac1}}

The representations $Q$ for $4$-strand braid carrying representation $R=[2]$ are
\begin{equation}
\begin{array}{r}
[2]^{\otimes 4}=[8]+3[7,1]+6[6,2]+3[6,1,1]+6[5,3]+8[5,2,1]+[5,1,1,1]+3[4,4]+7[4,3,1]+
\\
+6[4,2,2]+3[4,2,1,1]+3[3,3,2]+2[3,3,1,1]+3[3,2,2,1]+[2,2,2,2].
\end{array}
\end{equation}
Thus, there are five $3\times 3$ matrices, three $6\times 6$ matrices, one $7\times 7$ and one $8\times 8$ matrices. The eigenvalues in this case are $\lambda_1=q^6$, $\lambda_2=-q^2$ and $\lambda_3=1$. For the size $3\times 3$, the matrices are:
\begin{equation}
\begin{array}{l}
\mathcal{R}^{[7,1]}=\left(\begin{array}{crc}
q^6 \\ & q^6 \\ && -q^2
\end{array}\right),\ \ \ \
\mathcal{R}^{[6,1,1]}=\left(\begin{array}{ccc}
q^6 \\ & -q^2 \\ && -q^2
\end{array}\right),\ \ \ \
\mathcal{R}^{[4,4]}=\left(\begin{array}{ccc}
q^6 \\ & -q^2 \\ && 1
\end{array}\right),
\\
\mathcal{R}^{[4,2,1,1]}=\mathcal{R}^{[3,3,2]}=\left(\begin{array}{clc}
-q^2 \\ & -q^2 \\ && 1
\end{array}\right),\ \ \ \
\mathcal{R}^{[3,2,2,1]}=\left(\begin{array}{clc}
-q^2 \\ & 1 \\ && 1
\end{array}\right),
\mathcal{R}^{[3,3,1,1]}=\left(\begin{array}{clc}
-q^2   \\ && 1
\end{array}\right),
\end{array}
\end{equation}
\setlength{\arraycolsep}{6pt}
Then, the $U$-matrices are equal to
\begin{equation}
\begin{array}{l}
U^{[7,1]}\ \ =\left(\begin{array}{ccc}
1 \\
& \frac{[2]}{[4]} & \frac{\sqrt{[2][6]}}{[4]} \\
& \frac{\sqrt{[2][6]}}{[4]} & -\frac{[2]}{[4]}
\end{array}\right),\ \ \ \
U^{[6,1,1]}=\left(\begin{array}{ccc}
 \frac{[2]}{[4]} & \frac{\sqrt{[2][6]}}{[4]} \\
 \frac{\sqrt{[2][6]}}{[4]} & -\frac{[2]}{[4]} \\
 & & 1
\end{array}\right),
\\
U^{[4,4]}=\left(\begin{array}{ccc}
 -\frac{[2]}{[3][4]} & \frac{[2]\sqrt{[5]}}{[4]\sqrt{[3]}} & -\frac{\sqrt{[5]}}{[3]} \\
 \frac{[2]\sqrt{[5]}}{[4]\sqrt{[3]}} & -\frac{[6]}{[3][4]} & -\frac{1}{\sqrt{[3]}} \\
 -\frac{\sqrt{[5]}}{[3]} & -\frac{1}{\sqrt{[3]}} & -\frac{1}{[3]}
\end{array}\right),\ \ \ \
U^{[3,2,2,1]}\ \ =\left(\begin{array}{ccc}
 -\frac{1}{[2]} & -\frac{\sqrt{[3]}}{[2]} \\
 -\frac{\sqrt{[3]}}{[2]} & \frac{1}{[2]} \\
 & & 1
\end{array}\right),
\\
U^{[4,2,1,1]}=U^{[3,3,2]}=\left(\begin{array}{ccc}
1 \\
& \frac{1}{[2]} & \frac{\sqrt{[3]}}{[2]} \\
& \frac{\sqrt{[3]}}{[2]} & -\frac{1}{[2]}
\end{array}\right),

U^{[3,3,1,1]}=\left(\begin{array}{ccc}
& \frac{1}{[2]} & \frac{\sqrt{[3]}}{[2]} \\
& \frac{\sqrt{[3]}}{[2]} & -\frac{1}{[2]}
\end{array}\right),
\end{array}
\end{equation}
while the corresponding $W_1$ matrices are equal to
\begin{equation}
\begin{array}{l}
W_1^{[7,1]}\ \ =\left(\begin{array}{ccc}
\frac{[2]}{[6]} & \frac{\sqrt{[4][8]}}{[6]} \\
\frac{\sqrt{[4][8]}}{[6]} & -\frac{[2]}{[6]} \\
&& 1
\end{array}\right),\ \ \ \
W_1^{[6,1,1]}=\left(\begin{array}{ccc}
1 && \\
& -\frac{[2]}{[6]} & -\frac{\sqrt{[4][8]}}{[6]} \\
& -\frac{\sqrt{[4][8]}}{[6]} & \frac{[2]}{[6]} \\
\end{array}\right),\\
W_1^{[4,4]}=\left(\begin{array}{ccc}
1 && \\
& 1 & \\
&& 1 \\
\end{array}\right),\ \ \ \ \ \ \ \ \ \ \ \ \ \ \ \ \ \ \ \
W_1^{[3,2,2,1]}\ \ =\left(\begin{array}{ccc}
1 \\
& -\frac{1}{[3]} & \frac{\sqrt{[2][4]}}{[3]} \\
& \frac{\sqrt{[2][4]}}{[3]} & \frac{1}{[3]} \\
\end{array}\right),
\\
W_1^{[4,2,1,1]}=W^{[3,3,2]}=\left(\begin{array}{ccc}
\frac{1}{[3]} & \frac{\sqrt{[2][4]}}{[3]} \\
\frac{\sqrt{[2][4]}}{[3]} & -\frac{1}{[3]} \\
&& 1
\end{array}\right),
\\
W_1^{[3,3,1,1]}=\left(\begin{array}{ccc}
1 & 0 \\
0 & 1

\end{array}\right)
\end{array}
\end{equation}
which is in a full accordance with formulae (\ref{Um33}) and (\ref{Wm33}) from section \ref{s.eig3}.

Matrices having larger sizes are provided in Appendix \ref{app}.

\section{Conclusion\label{conc}}

In the present paper, we managed to present the generalization of the eigenvalue hypothesis suggested in \cite{IMMMec} for $3$-strands to a multi-strand case. We claim that, while the $3$-strand eigenvalue hypothesis comes from the Yang-Baxter equation, its generalization is based on solving the commutativity relations of non-neighbouring $\mathcal{R}$-matrices. In particular, in the case of four-strand braid, there is one commutativity relation. Solving it provided us with the eigenvalue answers for the $4$-strand Racah matrices of sizes up to $6\times 6$. Calculating matrices of the larger sizes encountered some computation difficulties, and, thus, they have not been found yet. However, all the eigenvalue formulae for the Racah matrices up to size $6\times 6$ has been checked by calculating the matrices for representations $[1]$ and $[2]$ using the highest weight method. The problem of finding the answers for matrices of larger sizes still remains.

Thus, the eigenvalue hypothesis for the particular example of the $4$-strand braids as well as the highest weight method allowed us to provide all the Racah matrices for the $4$-strand braids carrying representation $[2]$. This enabled to obtain the HOMFLY-PT polynomials for knots from 4-strand braids. Moreover, one can check the obtained results for the Racah matrices comparing known HOMFLY-PT polynomials in representation $[2]$ for various knots with those evaluated using the results of the present paper. We confirmed these results by comparing with correct answers for the HOMFLY-PT polynomials for: torus knots \cite{RJ}, twist knots $6_1$ and $7_2$ \cite{twist,db}, and many arborescent knots that have four-strand braid representation \cite{arborcalc,MMSRacah}.

In Appendix \ref{polex}, we list a few non-trivial examples of HOMFLY-PT polynomials in the first symmetric representation that have not been known so far: knots in accordance with the Rolfsen table \cite{katlas} which are non-arborescent and are described by four-strand braids: $9_{34}$, $9_{40}$, $9_{47}$, $9_{49}$ $10_{102}$, $10_{103}$, $10_{108}$, $10_{111}$, $10_{113}$, $10_{114}$, $10_{117}$, $10_{119}$, $10_{121}$, $10_{122}$, $10_{156}$, $10_{158}$, $10_{160}$, $10_{162}$-$10_{165}$.

The eigenvalue hypothesis in the multi-strand case was already related in \cite{ecA} to the well-established property \cite{IMMMfe}  $Al_{R}(q)=Al_{[1]}(q^{|R|})$ of the Alexander polynomials colored by the {\it single-hook} diagrams, which provided an indirect support for it. In this text, we found direct evidence in favor of this hypothesis and concrete formulae for its realization in the case of ${\cal R}$-matrices of small sizes. Extension to matrices of arbitrary size for any number of strands including three remains a challenging problem.



\section*{Acknowledgements}
Our work was partly supported by the grant of the Foundation for the Advancement of Theoretical Physics ``BASIS" (A.Mor. and A.S.), by RFBR grants 16-01-00291 (A.Mir.), 16-02-01021 (A.Mor.) and 17-01-00585 (An.Mor.), by joint grants 17-51-50051-YaF, 15-51-52031-NSC-a, 16-51-53034-GFEN, 16-51-45029-IND-a (A.M.'s), by grant 16-31-60082-mol-a-dk (A.S.). PR, VKS and SD acknowledge DST-RFBR grant(INT/RUS/RFBR/P-231) for support. SD would like to thank CSIR for research fellowship.

\appendix

\section{Representation [2] matrices\label{app}}

In this Appendix, all the Racah matrices of sizes larger than $3\times 3$ for the $4$-strand braid in the first symmetric representation are provided. The matrices of the size $6\times6$ are in full accordance with formulae (\ref{U66}) and (\ref{W66}) from section \ref{s.eig3}.

The diagonal $\mathcal{R}$-matrices look like
\begin{equation}
\begin{array}{l}
\mathcal{R}^{[6,2]}=\left(\begin{array}{clcccc}
q^6 \\& q^6 \\&&-q^2\\&&& q^6\\&&&&-q^2\\&&&&& 1
\end{array}\right),\ \ \ \
\mathcal{R}^{[5,3]}=\left(\begin{array}{clcccc}
-q^2 \\& -q^2 \\&& q^6\\&&& -q^2\\&&&& q^6\\&&&&& 1
\end{array}\right),\\
\mathcal{R}^{[4,2,2]}=\left(\begin{array}{clcccc}
1 \\& 1 \\&&-q^2\\&&& 1\\&&&&-q^2\\&&&&& q^6
\end{array}\right),
\mathcal{R}^{[4,3,1]}=\left(\begin{array}{clcccccc}
q^6 \\& -q^2 \\&&1\\&&& -q^2\\&&&& 1\\&&&&& -q^2 \\ &&&&&& -q^2
\end{array}\right),\\
\mathcal{R}^{[5,2,1]}=\left(\begin{array}{clccccccc}
-q^2 \\&- q^2 \\&& 1\\&&& q^6\\&&&& - q^2\\&&&&& q^6 \\ &&&&&& 1 \\ &&&&&&& -q^2
\end{array}\right).
\end{array}
\end{equation}

The $U$-matrices are equal to

\({U^{[6,2]}=\left(
\begin{array}{cccccc}
 1 & 0 & 0 & 0 & 0 & 0 \\
 0 & -\frac{[2]}{[4]} & \frac{\sqrt{[2] [6]}}{[4]} & 0 & 0 & 0 \\
 0 & -\frac{\sqrt{[2] [6]}}{[4]} & -\frac{[2]}{[4]} & 0 & 0 & 0 \\
 0 & 0 & 0 & \frac{[2]}{[3] [4]} & -\frac{[2]}{[4]} \sqrt{\frac{[5]}{[3]}} & \frac{\sqrt{[5]}}{[3]} \\
 0 & 0 & 0 & \frac{[2]}{[4]} \sqrt{\frac{[5]}{[3]}} & -\frac{[6]}{[3] [4]} & -\frac{1}{\sqrt{[3]}} \\
 0 & 0 & 0 & \frac{\sqrt{[5]}}{[3]} & \frac{1}{\sqrt{[3]}} & \frac{1}{[3]} \\
\end{array}
\right)}\),

\begin{equation}
\begin{array}{ll}
$\(U^{[4,2,2]}=\left(
\begin{array}{cccccc}
 1 & 0 & 0 & 0 & 0 & 0 \\
 0 & -\frac{1}{[2]} & \frac{\sqrt{[3]}}{[2]} & 0 & 0 & 0 \\
 0 & -\frac{\sqrt{[3]}}{[2]} & -\frac{1}{[2]} & 0 & 0 & 0 \\
 0 & 0 & 0 & \frac{1}{[3]} & -\frac{1}{\sqrt{[3]}} & -\frac{\sqrt{[5]}}{[3]} \\
 0 & 0 & 0 & \frac{1}{\sqrt{[3]}} & -\frac{[6]}{[3] [4]} & \frac{[2]}{[4]} \sqrt{\frac{[5]}{[3]}} \\
 0 & 0 & 0 & -\frac{\sqrt{[5]}}{[3]} & -\frac{[2]}{[4]} \sqrt{\frac{[5]}{[3]}} & \frac{[2]}{[3] [4]} \\
\end{array}
\right)\)$,\\
$\(~{U^{[5,3]}=\left(
\begin{array}{cccccc}
 1 & 0 & 0 & 0 & 0 & 0 \\
 0 & -\frac{[2]}{[4]} & -\frac{\sqrt{[2] [6]}}{[4]} & 0 & 0 & 0 \\
 0 & \frac{\sqrt{[2] [6]}}{[4]} & -\frac{[2]}{[4]} & 0 & 0 & 0 \\
 0 & 0 & 0 & \frac{-[6]}{[4][3]} & \frac{[2]}{[4]} \sqrt{\frac{[5]}{[3]}} & -\frac{1}{\sqrt{[3]}} \\
 0 & 0 & 0 & -\frac{[2]}{[4]} \sqrt{\frac{[5]}{[3]}} & \frac{[2]}{[3][4]} & \frac{\sqrt{[5]}}{[3]} \\
 0 & 0 & 0 & \frac{1}{\sqrt{[3]}} & \frac{\sqrt{[5]}}{[3]} & \frac{1}{[3]} \\
\end{array}
\right)}\)$,\\

$\(~U^{[4,3,1]}\text{=}\left(
\begin{array}{ccccccc}
 \frac{[2]}{[3][4]} & \frac{[2] [5]}{[4] \sqrt{[3] [5]}} & \frac{\sqrt{[5]}}{[3]} & 0 & 0 & 0 & 0 \\
 \frac{[2] [5]}{[4] \sqrt{[3][5]}} & \frac{[6]}{[3][4]} & -\frac{1}{\sqrt{[3]}} & 0 & 0 & 0 & 0 \\
 \frac{\sqrt{[5]}}{[3]} & -\frac{1}{\sqrt{[3]}} & \frac{1}{[3]} & 0 & 0 & 0 & 0 \\
 0 & 0 & 0 & \frac{1}{[2]} & \frac{\sqrt{[3]}}{[2]} & 0 & 0 \\
 0 & 0 & 0 & \frac{\sqrt{[3]}}{[2]} & -\frac{1}{[2]} & 0 & 0 \\
 0 & 0 & 0 & 0 & 0 & 1 & 0 \\
 0 & 0 & 0 & 0 & 0 & 0 & 1 \\
\end{array}
\right)\)$,\\
$\(~{U^{[5,2,1]}\text{=}\left(
\begin{array}{cccccccc}
 1 & 0 & 0 & 0 & 0 & 0 & 0 & 0 \\
 0 & \frac{1}{[2]} & \frac{\sqrt{[3]}}{[2]} & 0 & 0 & 0 & 0 & 0 \\
 0 & \frac{\sqrt{[3]}}{[2]} & \frac{-1}{[2]}&0 & 0 & 0 & 0 & 0 \\
 0 & 0 & 0 & \frac{[2]}{[4]} & \frac{\sqrt{[2] [6]}}{[4]} & 0 & 0 & 0 \\
 0 & 0 & 0 & \frac{\sqrt{[2] [6]}}{[4]} & -\frac{[2]}{[4]} & 0 & 0 & 0 \\
 0 & 0 & 0 & 0 & 0 & \frac{[2]}{[3][4]} & \frac{\sqrt{[5]}}{[3]} & -\frac{[2] [5]}{[4] \sqrt{[3][5]}} \\
 0 & 0 & 0 & 0 & 0 & \frac{\sqrt{[5]}}{[3]} & \frac{1}{[3]} & \frac{1}{\sqrt{[3]}} \\
 0 & 0 & 0 & 0 & 0 & \frac{[2] [5]}{[4] \sqrt{[3][5]}} & -\frac{1}{\sqrt{[3]}} & -\frac{[6]}{[3][4]} \\
\end{array}
\right)}\)$.
\end{array}
\end{equation}
The $W_1$-matrices are equal to
\begin{equation}
\begin{array}{ll}
$\(~{W_1^{[6,2]}=\left(
\begin{array}{cccccc}
 \frac{[2]}{[5] [6]} & \frac{[2] \sqrt{\frac{[7]}{[5]}}}{[6]} & 0 & \frac{\sqrt{[3] [7]}}{[5]} & 0 & 0 \\
 \frac{[2] \sqrt{\frac{[7]}{[5]}}}{[6]} & 1-\frac{[2]^2}{[4] [6]} & 0 & -\frac{[2]}{[4]} \sqrt{\frac{[3]}{[5]}} & 0 & 0 \\
 0 & 0 & -\frac{[2]}{[4]} & 0 & -\frac{\sqrt{[2] [6]}}{[4]} & 0 \\
 \frac{\sqrt{[3] [7]}}{[5]} & -\frac{[2]}{[4]} \sqrt{\frac{[3]}{[5]}} & 0 & \frac{[2]}{[4] [5]} & 0 & 0 \\
 0 & 0 & -\frac{\sqrt{[2] [6]}}{[4]} & 0 & \frac{[2]}{[4]} & 0 \\
 0 & 0 & 0 & 0 & 0 & 1 \\
\end{array}
\right)}\)$,\\
$\(W_1^{[4,2,2]}=\left(
\begin{array}{cccccc}
 \frac{[2]}{[3] [4]} & \frac{[2]}{[3] \sqrt{\frac{[4]}{[2]}}} & 0 & \frac{\sqrt{[2] [6]}}{[4]} & 0 & 0 \\
 \frac{[2]}{[3] \sqrt{\frac{[4]}{[2]}}} & \frac{-1+[5]+[7]}{[3] [5]} & 0 & -\frac{[2]}{[5]} \sqrt{\frac{[6]}{[4]}} & 0 & 0 \\
 0 & 0 & -\frac{1}{[5]} & 0 & -\frac{\sqrt{[4] [6]}}{[5]} & 0 \\
 \frac{\sqrt{[2] [6]}}{[4]} & -\frac{[2]}{[5]} \sqrt{\frac{[6]}{[4]}} & 0 & \frac{[2]}{[4] [5]} & 0 & 0 \\
 0 & 0 & -\frac{\sqrt{[4] [6]}}{[5]} & 0 & \frac{1}{[5]} & 0 \\
 0 & 0 & 0 & 0 & 0 & 1 \\
\end{array}
\right)
\)$,\\
$\(~W_1^{[5,3]}=\left(
\begin{array}{cccccc}
 \frac{1}{[3]} & \frac{\sqrt{[5]}}{[3]} & 0 & -\frac{1}{\sqrt{[3]}} & 0 & 0 \\
 \frac{\sqrt{[5]}}{[3]} & \frac{[2]}{[4][3]} & 0 & \frac{[2]}{[4]} \sqrt{\frac{[5]}{[3]}} & 0 & 0 \\
 0 & 0 & -\frac{[2]}{[4]} & 0 & -\frac{\sqrt{[2] [6]}}{[4]} & 0 \\
 -\frac{1}{\sqrt{[3]}} & \frac{[2]}{[4]} \sqrt{\frac{[5]}{[3]}} & 0 & \frac{[6]}{[4][3]} & 0 & 0 \\
 0 & 0 & -\frac{\sqrt{[2] [6]}}{[4]} & 0 & \frac{[2]}{[4]} & 0 \\
 0 & 0 & 0 & 0 & 0 & -1 \\
\end{array}
\right)\)$,\\

$\(W_1^{[5,2,1]}=
\left(
{\tiny\begin{array}{cccccccc}
 \frac{[8]}{[4] [6]} & -\frac{\sqrt{[2]}}{\sqrt{[3] [6]}} & 0 & 0 & -\frac{\sqrt{[7]}}{[6]} & 0 & 0 & \frac{\sqrt{[7]}}{\sqrt{[2] [6]}} \\
 \frac{\sqrt{[2]}}{\sqrt{[3] [6]}} & -\frac{[2]}{[3] [5]} & 0 & 0 & \frac{\sqrt{[2] [7]}}{\sqrt{[3] [6]}} & 0 & 0 & \frac{[7]}{[5] \sqrt{[3] [7]}}
\\
 0 & 0 & \frac{[2]}{[5]} & 0 & 0 & 0 & -\frac{\sqrt{[3] [7]}}{[5]} & 0 \\
 0 & 0 & 0 & \frac{1}{[4]} & 0 & -\frac{\sqrt{[3] [5]}}{[4]} & 0 & 0 \\
 \frac{\sqrt{[7]}}{[6]} & \frac{\sqrt{[2] [7]}}{\sqrt{[3] [6]}} & 0 & 0 & -\frac{[2]}{[4] [6]} & 0 & 0 & \frac{\sqrt{[2]}}{[4] \sqrt{[6]}} \\
 0 & 0 & 0 & -\frac{\sqrt{[3] [5]}}{[4]} & 0 & -\frac{1}{[4]} & 0 & 0 \\
 0 & 0 & -\frac{\sqrt{[3] [7]}}{[5]} & 0 & 0 & 0 & -\frac{[2]}{[5]} & 0 \\
 \frac{\sqrt{[7]}}{\sqrt{[2] [6]}} & -\frac{[7]}{[5] \sqrt{[3] [7]}} & 0 & 0 & -\frac{\sqrt{[2]}}{[4] \sqrt{[6]}} & 0 & 0 & -\frac{[1]+[3]+[7]}{[4]
[5]} \\
\end{array}}
\right)\)$,\\

$\(~{W_1^{[4,3,1]}\text{=}\left(
\begin{array}{ccccccc}
 1 & 0 & 0 & 0 & 0 & 0 & 0 \\
 0 & \frac{[6]}{[3][2] [5]} & 0 & \frac{[4] \sqrt{[4][6]}}{[2][3] [5]} & 0 & \frac{[4]}{[2] \sqrt{[3][5]}} & \frac{[4]}{[2] \sqrt{[3][5]}}
\\
 0 & 0 & -\frac{1}{[5]} & 0 & \frac{\sqrt{[4][6]}}{[5]} & 0 & 0 \\
 0 & \frac{[4] \sqrt{[4][6]}}{[2][3] [5]} & 0 & \frac{[4]+[2] [5]}{[2] [3] [5]} & 0 &-\frac{\sqrt{\frac{[4][6]}{[3][5]}}}{[3] } & -\frac{\sqrt{\frac{[4][6]}{[3][5]}}}{[3] } \\
 0 & 0 & \frac{\sqrt{[4][6]}}{[5]} & 0 & \frac{1}{[5]} & 0 & 0 \\
 0 & \frac{[4]}{[2] \sqrt{[3][5]}} & 0 & -\frac{\sqrt{\frac{[4][6]}{[3][5]}}}{[3] } & 0 & -\frac{1}{[3]} & \frac{[4]}{[2]
[3]} \\
 0 & \frac{[4]}{[2] \sqrt{[3][5]}} & 0 & -\frac{\sqrt{\frac{[4][6]}{[3][5]}}}{[3] } & 0 & \frac{[4]}{[2] [3]} & -\frac{1}{[3]}
\\
\end{array}
\right)}\)$.
\end{array}
\end{equation}

\section{Polynomial examples\label{polex}}

In this appendix four-strand HOMFLY polynomials in representation [2] are listed.  We calculated these polynomials using 
the matrices from the Appendix \ref{app} and present them in the matrix form suggested in \cite{twist}. The matrix describes
the coefficients of a polynomial in $A^2$ and $q^2$ as

\setlength{\arraycolsep}{1pt}

\begin{equation}
\nonumber q^{10}A^{16} \left(\begin{array}{rr}
3 & 4 \\
& \\
1 & 2 \\
\end{array}\right)
= q^{10}A^{16}+2q^{12}A^{16}+3q^{10}A^{18}+4q^{12}A^{18}.
\end{equation}

HOMFLY polynomials for knots $6_1$ and $7_2$ are already known since those are arborescent knots, and the answers calculated using four-strand representations coincide with those. All other polynomials presented here were unknown before since those are non-arborescent knots. Their limit for $A=q^2$, colored Jones polynomials, are known and are correct. Other limits (Alexander $A=1$ and special $q=1$ polynomials) are known because their dependence on the representation is quite simple \cite{DMMSS,Cab}:
\begin{equation}
\begin{array}{lcl}
\mathcal{A}_{V}(q)=H_{V}(A=1,q),& \ \ & \mathcal{A}_{V}(q)=P_{[1]}(q^{|V|}), V\ -\ \text{hook diagram}\\
\\
\sigma_{V}(A)=H_{V}(A,q=1),& \ \ & \sigma_{V}(A)=(\sigma_{[1]}(A))^{|V|}.
\end{array}
\end{equation}
These limits of the polynomials below are also correct.

\begin{equation*}
H_{[2]}^{6_1}=\frac{1}{q^{10}A^{8}}
\left(\begin{array}{rrrrrrrrr}
0 & 0 & 0 & 0 & 0 & 0 & 0 & 1 & 0 \\
&&&&&&&& \\
0 & 0 & 0 & 0 & -1 & 0 & 2 & 0 & -1 \\
&&&&&&&& \\
0 & 0 & 1 & -2 & -2 & 4 & 0 & -3 & 0 \\
&&&&&&&& \\
0 & 1 & -1 & -2 & 3 & 2 & -2 & 0 & 1 \\
&&&&&&&& \\
1 & -1 & -1 & 2 & 1 & -1 & 0 & 0 & 0 \\
&&&&&&&& \\
-1 & -1 & 1 & 0 & -1 & 0 & 0 & 0 & 0 \\
&&&&&&&& \\
0 & 1 & 0 & 0 & 0 & 0 & 0 & 0 & 0
\end{array}\right)
\ \ \ \ \ H_{[2]}^{7_2}=\frac{1}{q^{10}A^{8}}
\left(\begin{array}{rrrrrrrrr}
0 & 0 & 0 & 0 & 0 & 0 & 0 & 1 & 0 \\
&&&&&&&& \\
0 & 0 & 0 & 0 & -1 & 0 & 2 & 0 & -1 \\
&&&&&&&& \\
0 & 0 & 1 & -2 & -2 & 4 & 0 & -3 & 0 \\
&&&&&&&& \\
0 & 1 & -1 & -2 & 3 & 2 & -2 & 0 & 1 \\
&&&&&&&& \\
1 & -1 & -1 & 2 & 1 & -1 & 0 & 0 & 0 \\
&&&&&&&& \\
-1 & -1 & 1 & 0 & -1 & 0 & 0 & 0 & 0 \\
&&&&&&&& \\
0 & 1 & 0 & 0 & 0 & 0 & 0 & 0 & 0
\end{array}\right)
\end{equation*}
\begin{equation*}
H_{[2]}^{9_{34}}=\frac{1}{q^{18}A^{8}}
\left(\begin{array}{rrrrrrrrrrrrrrrrrrr}
0 & 0 & 0 & 0 & 0 & 1 & -3 & 0 & 10 & -10 & -9 & 21 & -3 & -14 & 8 & 2 & -3 & 1 & 0 \\
&&&&&&&&&&&&&&&&&& \\
0 & 0 & -1 & 2 & 3 & -12 & 3 & 28 & -31 & -24 & 55 & -6 & -41 & 21 & 11 & -12 & 1 & 2 & -1 \\
&&&&&&&&&&&&&&&&&& \\
1 & -3 & 2 & 9 & -21 & 0 & 50 & -44 & -44 & 84 & -3 & -65 & 34 & 18 & -22 & 5 & 4 & -3 & 1 \\
&&&&&&&&&&&&&&&&&& \\
-2 & 3 & 7 & -20 & 1 & 47 & -38 & -46 & 74 & 5 & -62 & 26 & 19 & -20 & 2 & 4 & -2 & 0 & 0 \\
&&&&&&&&&&&&&&&&&& \\
1 & 2 & -11 & 2 & 28 & -25 & -29 & 48 & 3 & -37 & 16 & 10 & -9 & 1 & 1 & 0 & 0 & 0 & 0 \\
&&&&&&&&&&&&&&&&&& \\
0 & -2 & 2 & 8 & -10 & -9 & 18 & 0 & -13 & 5 & 3 & -2 & 0 & 0 & 0 & 0 & 0 & 0 & 0 \\
&&&&&&&&&&&&&&&&&& \\
0 & 0 & 1 & -2 & -1 & 4 & -1 & -2 & 1 & 0 & 0 & 0 & 0 & 0 & 0 & 0 & 0 & 0 & 0
\end{array}\right)
\end{equation*}
\begin{equation*}
H_{[2]}^{9_{40}}=\frac{1}{q^{16}}
\left(\begin{array}{rrrrrrrrrrrrrrrrrrr}
0 & 0 & 0 & 0 & 0 & 0 & 0 & 0 & 0 & 0 & 1 & -2 & -1 & 4 & -1 & -2 & 1 & 0 & 0 \\
&&&&&&&&&&&&&&&&&& \\
0 & 0 & 0 & 0 & 0 & 0 & 0 & -2 & 2 & 8 & -10 & -10 & 19 & 1 & -15 & 5 & 4 & -2 & 0 \\
&&&&&&&&&&&&&&&&&& \\
0 & 0 & 0 & 0 & 1 & 2 & -12 & 3 & 33 & -31 & -33 & 59 & 1 & -45 & 21 & 12 & -11 & 0 & 1 \\
&&&&&&&&&&&&&&&&&& \\
0 & 0 & -2 & 4 & 7 & -25 & 2 & 57 & -46 & -56 & 86 & 4 & -72 & 33 & 22 & -25 & 3 & 6 & -2 \\
&&&&&&&&&&&&&&&&&& \\
1 & -4 & 3 & 13 & -28 & -4 & 66 & -39 & -62 & 87 & 10 & -72 & 36 & 21 & -28 & 7 & 4 & -4 & 1 \\
&&&&&&&&&&&&&&&&&& \\
-1 & 3 & 4 & -17 & -1 & 37 & -25 & -41 & 50 & 7 & -46 & 19 & 14 & -15 & 2 & 3 & -1 & 0 & 0 \\
&&&&&&&&&&&&&&&&&& \\
0 & 1 & -4 & 0 & 14 & -10 & -13 & 22 & 0 & -15 & 10 & 2 & -4 & 1 & 0 & 0 & 0 & 0 & 0
\end{array}\right)
\end{equation*}
\begin{equation*}
H_{[2]}^{9_{47}}=\frac{1}{q^{16}}
\left(\begin{array}{rrrrrrrrrrrrrrrrrrr}
0 & 0 & 0 & 0 & 0 & 0 & 0 & 0 & 0 & 0 & 0 & 0 & 0 & 0 & 1 & 0 & 0 & 0 & 0 \\
&&&&&&&&&&&&&&&&&& \\
0 & 0 & 0 & 0 & 0 & 0 & 0 & 0 & 0 & -1 & 1 & 0 & -3 & 1 & 1 & -2 & -1 & 0 & 0 \\
&&&&&&&&&&&&&&&&&& \\
0 & 0 & 0 & 0 & 0 & 1 & -1 & -1 & 3 & 1 & -4 & 2 & 3 & -1 & 1 & 0 & 0 & 2 & 0 \\
&&&&&&&&&&&&&&&&&& \\
0 & 0 & -1 & 1 & 2 & -4 & -2 & 8 & 0 & -10 & 6 & 6 & -7 & 1 & 1 & -3 & 1 & 0 & -1 \\
&&&&&&&&&&&&&&&&&& \\
1 & -2 & 0 & 5 & -6 & -6 & 12 & 0 & -15 & 7 & 7 & -10 & 3 & 2 & -5 & 4 & 1 & -2 & 1 \\
&&&&&&&&&&&&&&&&&& \\
-1 & 1 & 4 & -4 & -4 & 9 & 1 & -10 & 4 & 6 & -6 & 2 & 2 & -4 & 2 & 1 & -1 & 0 & 0 \\
&&&&&&&&&&&&&&&&&& \\
0 & 1 & -2 & -2 & 4 & 0 & -4 & 3 & 3 & -3 & 2 & 0 & -2 & 1 & 0 & 0 & 0 & 0 & 0
\end{array}\right)
\end{equation*}
\begin{equation*}
H_{[2]}^{9_{49}}=\frac{A^{8}}{q^{8}}
\left(\begin{array}{rrrrrrrrrrrrrrr}
0 & 0 & 0 & 0 & 0 & 0 & 0 & 0 & 0 & 0 & -1 & 0 & 1 & 0 & 0 \\
&&&&&&&&&&&&&& \\
0 & 0 & 0 & 0 & 0 & 0 & 1 & 4 & -1 & -2 & 5 & 2 & -2 & 2 & 0 \\
&&&&&&&&&&&&&& \\
0 & 0 & 0 & -1 & -4 & 0 & 4 & -9 & -8 & 7 & -3 & -8 & 2 & -1 & -3 \\
&&&&&&&&&&&&&& \\
0 & 1 & 1 & -5 & 3 & 12 & -7 & -7 & 16 & 0 & -7 & 8 & 1 & -3 & 3 \\
&&&&&&&&&&&&&& \\
2 & 1 & -6 & 1 & 10 & -7 & -9 & 11 & 0 & -8 & 6 & 0 & -3 & 2 & 0 \\
&&&&&&&&&&&&&& \\
1 & -2 & -1 & 5 & -2 & -5 & 5 & 1 & -4 & 3 & 0 & -2 & 1 & 0 & 0
\end{array}\right)
\end{equation*}
\begin{equation*}
H_{[2]}^{10_{102}}=\frac{1}{q^{18}A^{4}}
\left(\begin{array}{rrrrrrrrrrrrrrrrrrrrr}
0 & 0 & 0 & 0 & 0 & 0 & 0 & 1 & -2 & 0 & 5 & -6 & -2 & 10 & -5 & -5 & 6 & 0 & -2 & 1 & 0 \\
&&&&&&&&&&&&&&&&&&&& \\
0 & 0 & 0 & 0 & -1 & 2 & 1 & -8 & 7 & 8 & -21 & 6 & 22 & -20 & -9 & 18 & -2 & -8 & 3 & 1 & -1 \\
&&&&&&&&&&&&&&&&&&&& \\
0 & 0 & 0 & -2 & 4 & 2 & -14 & 15 & 12 & -39 & 16 & 39 & -41 & -13 & 37 & -9 & -15 & 10 & 1 & -3 & 1 \\
&&&&&&&&&&&&&&&&&&&& \\
1 & -1 & -3 & 8 & -1 & -19 & 25 & 12 & -54 & 24 & 50 & -53 & -14 & 44 & -11 & -16 & 11 & 1 & -3 & 1 & 0 \\
&&&&&&&&&&&&&&&&&&&& \\
0 & -2 & 4 & 0 & -15 & 18 & 15 & -48 & 13 & 51 & -43 & -19 & 38 & -7 & -15 & 9 & 1 & -3 & 1 & 0 & 0 \\
&&&&&&&&&&&&&&&&&&&& \\
-1 & 2 & 1 & -8 & 8 & 10 & -24 & 3 & 27 & -18 & -12 & 17 & -1 & -7 & 3 & 1 & -1 & 0 & 0 & 0 & 0 \\
&&&&&&&&&&&&&&&&&&&& \\
0 & 1 & -2 & 0 & 5 & -5 & -3 & 9 & -3 & -5 & 5 & 0 & -2 & 1 & 0 & 0 & 0 & 0 & 0 & 0 & 0
\end{array}\right)
\end{equation*}
\begin{equation*}
H_{[2]}^{10_{103}}=\frac{1}{q^{24}A^{12}}
\left(\begin{array}{rrrrrrrrrrrrrrrrrrrrr}
0 & 0 & 0 & 0 & 0 & 0 & 0 & 1 & -2 & 1 & 4 & -6 & 0 & 8 & -6 & -3 & 6 & -1 & -2 & 1 & 0 \\
&&&&&&&&&&&&&&&&&&&& \\
0 & 0 & 0 & 0 & -1 & 1 & 2 & -8 & 3 & 12 & -20 & -2 & 27 & -17 & -17 & 20 & 2 & -11 & 2 & 2 & -1 \\
&&&&&&&&&&&&&&&&&&&& \\
0 & 0 & 1 & -3 & 2 & 8 & -17 & 5 & 31 & -41 & -6 & 61 & -31 & -37 & 43 & 7 & -23 & 5 & 6 & -2 & 0 \\
&&&&&&&&&&&&&&&&&&&& \\
0 & 1 & -3 & 2 & 9 & -19 & 5 & 37 & -46 & -17 & 73 & -28 & -52 & 46 & 14 & -29 & 4 & 9 & -4 & -1 & 1 \\
&&&&&&&&&&&&&&&&&&&& \\
1 & -3 & 2 & 7 & -17 & 4 & 30 & -38 & -16 & 59 & -21 & -41 & 33 & 8 & -19 & 3 & 4 & -2 & 0 & 0 & 0 \\
&&&&&&&&&&&&&&&&&&&& \\
-1 & 1 & 2 & -7 & 3 & 13 & -19 & -6 & 29 & -11 & -19 & 17 & 4 & -9 & 2 & 2 & -1 & 0 & 0 & 0 & 0 \\
&&&&&&&&&&&&&&&&&&&& \\
0 & 1 & -2 & 1 & 4 & -7 & 1 & 9 & -7 & -3 & 6 & -1 & -2 & 1 & 0 & 0 & 0 & 0 & 0 & 0 & 0
\end{array}\right)
\end{equation*}
\begin{equation*}
H_{[2]}^{10_{108}}=\frac{1}{q^{20}A^{4}}
\left(\begin{array}{rrrrrrrrrrrrrrrrrrrrr}
0 & 0 & 0 & 0 & 0 & 0 & 0 & 1 & -2 & -1 & 5 & -2 & -5 & 5 & 1 & -4 & 3 & 0 & -2 & 1 & 0 \\
&&&&&&&&&&&&&&&&&&&& \\
0 & 0 & 0 & 0 & -1 & 1 & 4 & -6 & -6 & 14 & 3 & -19 & 5 & 14 & -11 & -1 & 6 & -5 & 1 & 2 & -1 \\
&&&&&&&&&&&&&&&&&&&& \\
0 & 0 & 1 & -3 & 0 & 11 & -9 & -17 & 25 & 12 & -35 & 5 & 27 & -17 & -5 & 11 & -7 & 0 & 3 & -2 & 0 \\
&&&&&&&&&&&&&&&&&&&& \\
0 & 1 & -3 & -1 & 12 & -7 & -21 & 23 & 17 & -36 & 1 & 29 & -17 & -7 & 15 & -8 & -3 & 7 & -2 & -1 & 1 \\
&&&&&&&&&&&&&&&&&&&& \\
1 & -3 & -1 & 11 & -7 & -17 & 21 & 10 & -28 & 6 & 19 & -15 & 0 & 10 & -8 & 0 & 4 & -2 & 0 & 0 & 0 \\
&&&&&&&&&&&&&&&&&&&& \\
-1 & 1 & 5 & -5 & -7 & 13 & 3 & -16 & 6 & 10 & -11 & 1 & 6 & -6 & 0 & 2 & -1 & 0 & 0 & 0 & 0 \\
&&&&&&&&&&&&&&&&&&&& \\
0 & 1 & -2 & -2 & 5 & -1 & -6 & 5 & 2 & -5 & 3 & 1 & -2 & 1 & 0 & 0 & 0 & 0 & 0 & 0 & 0
\end{array}\right)
\end{equation*}
\begin{equation*}
H_{[2]}^{10_{111}}=\frac{A^{4}}{q^{14}}
\left(\begin{array}{rrrrrrrrrrrrrrrrrrrrr}
0 & 0 & 0 & 0 & 0 & 0 & 0 & 1 & -2 & -1 & 6 & -1 & -7 & 5 & 3 & -5 & 2 & 1 & -2 & 1 & 0 \\
&&&&&&&&&&&&&&&&&&&& \\
0 & 0 & 0 & 0 & -1 & 2 & 3 & -11 & -3 & 22 & -8 & -26 & 20 & 10 & -21 & 6 & 6 & -7 & 2 & 1 & -1 \\
&&&&&&&&&&&&&&&&&&&& \\
0 & 0 & 0 & -3 & 5 & 10 & -22 & -9 & 49 & -12 & -51 & 45 & 22 & -43 & 16 & 16 & -16 & 4 & 4 & -3 & 1 \\
&&&&&&&&&&&&&&&&&&&& \\
1 & 0 & -6 & 7 & 17 & -29 & -17 & 59 & -11 & -65 & 47 & 25 & -55 & 15 & 20 & -24 & 4 & 6 & -5 & 1 & 0 \\
&&&&&&&&&&&&&&&&&&&& \\
0 & -5 & 3 & 15 & -22 & -18 & 46 & -5 & -52 & 36 & 22 & -42 & 14 & 18 & -19 & 5 & 5 & -4 & 1 & 0 & 0 \\
&&&&&&&&&&&&&&&&&&&& \\
-1 & 2 & 5 & -9 & -5 & 21 & -4 & -22 & 19 & 9 & -20 & 7 & 8 & -8 & 1 & 2 & -1 & 0 & 0 & 0 & 0 \\
&&&&&&&&&&&&&&&&&&&& \\
0 & 1 & -2 & -1 & 6 & -3 & -6 & 8 & 0 & -6 & 4 & 1 & -2 & 1 & 0 & 0 & 0 & 0 & 0 & 0 & 0
\end{array}\right)
\end{equation*}
\begin{equation*}
H_{[2]}^{10_{113}}=\frac{1}{q^{16}}
\left(\begin{array}{rrrrrrrrrrrrrrrrrrrrr}
0 & 0 & 0 & 0 & 0 & 0 & 0 & 1 & -4 & 1 & 13 & -15 & -10 & 27 & -6 & -17 & 12 & 2 & -4 & 1 & 0 \\
&&&&&&&&&&&&&&&&&&&& \\
0 & 0 & 0 & 0 & -1 & 4 & 2 & -21 & 15 & 38 & -61 & -19 & 83 & -29 & -51 & 40 & 8 & -19 & 3 & 3 & -1 \\
&&&&&&&&&&&&&&&&&&&& \\
0 & 0 & 0 & -4 & 9 & 7 & -44 & 30 & 77 & -108 & -32 & 152 & -52 & -89 & 82 & 11 & -41 & 15 & 6 & -5 & 1 \\
&&&&&&&&&&&&&&&&&&&& \\
1 & -1 & -7 & 16 & 7 & -62 & 36 & 94 & -132 & -48 & 172 & -59 & -107 & 92 & 10 & -48 & 17 & 6 & -6 & 1 & 0 \\
&&&&&&&&&&&&&&&&&&&& \\
0 & -3 & 10 & 6 & -44 & 23 & 81 & -94 & -49 & 139 & -33 & -88 & 71 & 12 & -36 & 12 & 5 & -4 & 1 & 0 & 0 \\
&&&&&&&&&&&&&&&&&&&& \\
-1 & 3 & 2 & -18 & 9 & 36 & -41 & -25 & 62 & -12 & -40 & 28 & 7 & -14 & 3 & 2 & -1 & 0 & 0 & 0 & 0 \\
&&&&&&&&&&&&&&&&&&&& \\
0 & 1 & -3 & 0 & 9 & -8 & -8 & 15 & -2 & -10 & 7 & 1 & -3 & 1 & 0 & 0 & 0 & 0 & 0 & 0 & 0
\end{array}\right)
\end{equation*}
\begin{equation*}
H_{[2]}^{10_{114}}=\frac{1}{q^{22}A^{8}}
\left(\begin{array}{rrrrrrrrrrrrrrrrrrrrr}
0 & 0 & 0 & 0 & 0 & 0 & 0 & 1 & -3 & 0 & 9 & -9 & -7 & 17 & -4 & -11 & 8 & 1 & -3 & 1 & 0 \\
&&&&&&&&&&&&&&&&&&&& \\
0 & 0 & 0 & 0 & -1 & 2 & 4 & -12 & -1 & 30 & -22 & -33 & 49 & 6 & -44 & 17 & 16 & -13 & 0 & 3 & -1 \\
&&&&&&&&&&&&&&&&&&&& \\
0 & 0 & 1 & -4 & 2 & 14 & -25 & -10 & 65 & -34 & -71 & 88 & 20 & -83 & 27 & 32 & -26 & -1 & 7 & -2 & 0 \\
&&&&&&&&&&&&&&&&&&&& \\
0 & 1 & -4 & 2 & 16 & -27 & -16 & 72 & -29 & -86 & 88 & 35 & -92 & 24 & 43 & -32 & -3 & 13 & -4 & -2 & 1 \\
&&&&&&&&&&&&&&&&&&&& \\
1 & -4 & 2 & 14 & -24 & -11 & 60 & -24 & -65 & 70 & 23 & -67 & 21 & 26 & -23 & 1 & 6 & -2 & 0 & 0 & 0 \\
&&&&&&&&&&&&&&&&&&&& \\
-1 & 2 & 4 & -12 & -2 & 28 & -16 & -32 & 36 & 8 & -35 & 13 & 12 & -12 & 1 & 3 & -1 & 0 & 0 & 0 & 0 \\
&&&&&&&&&&&&&&&&&&&& \\
0 & 1 & -3 & 0 & 9 & -8 & -7 & 15 & -3 & -9 & 8 & 0 & -3 & 1 & 0 & 0 & 0 & 0 & 0 & 0 & 0
\end{array}\right)
\end{equation*}
\begin{equation*}
H_{[2]}^{10_{117}}=\frac{1}{q^{24}A^{12}}
\left(\begin{array}{rrrrrrrrrrrrrrrrrrrrr}
0 & 0 & 0 & 0 & 0 & 0 & 0 & 1 & -3 & 1 & 7 & -10 & -2 & 15 & -8 & -8 & 9 & 0 & -3 & 1 & 0 \\
&&&&&&&&&&&&&&&&&&&& \\
0 & 0 & 0 & 0 & -1 & 2 & 3 & -13 & 7 & 25 & -38 & -9 & 57 & -25 & -37 & 34 & 8 & -17 & 2 & 3 & -1 \\
&&&&&&&&&&&&&&&&&&&& \\
0 & 0 & 1 & -4 & 4 & 11 & -32 & 12 & 60 & -81 & -24 & 122 & -50 & -81 & 73 & 18 & -40 & 6 & 9 & -3 & 0 \\
&&&&&&&&&&&&&&&&&&&& \\
0 & 1 & -5 & 5 & 14 & -40 & 12 & 76 & -94 & -41 & 150 & -46 & -108 & 85 & 30 & -52 & 7 & 14 & -6 & -1 & 1 \\
&&&&&&&&&&&&&&&&&&&& \\
1 & -4 & 5 & 11 & -33 & 12 & 64 & -79 & -37 & 125 & -35 & -89 & 64 & 21 & -36 & 5 & 7 & -3 & 0 & 0 & 0 \\
&&&&&&&&&&&&&&&&&&&& \\
-1 & 2 & 2 & -14 & 8 & 29 & -43 & -18 & 67 & -19 & -46 & 32 & 11 & -16 & 2 & 3 & -1 & 0 & 0 & 0 & 0 \\
&&&&&&&&&&&&&&&&&&&& \\
0 & 1 & -3 & 2 & 8 & -14 & -2 & 21 & -10 & -10 & 10 & 0 & -3 & 1 & 0 & 0 & 0 & 0 & 0 & 0 & 0
\end{array}\right)
\end{equation*}
\begin{equation*}
H_{[2]}^{10_{121}}=\frac{1}{q^{16}}
\left(\begin{array}{rrrrrrrrrrrrrrrrrrrrr}
0 & 0 & 0 & 0 & 0 & 0 & 0 & 1 & -3 & 0 & 10 & -9 & -11 & 20 & 0 & -14 & 7 & 2 & -3 & 1 & 0 \\
&&&&&&&&&&&&&&&&&&&& \\
0 & 0 & 0 & 0 & -1 & 3 & 2 & -17 & 10 & 36 & -48 & -28 & 74 & -13 & -51 & 29 & 11 & -14 & 2 & 2 & -1 \\
&&&&&&&&&&&&&&&&&&&& \\
0 & 0 & 0 & -3 & 8 & 6 & -40 & 22 & 80 & -99 & -52 & 151 & -31 & -101 & 70 & 21 & -38 & 9 & 6 & -4 & 1 \\
&&&&&&&&&&&&&&&&&&&& \\
1 & -1 & -7 & 15 & 10 & -63 & 29 & 109 & -128 & -72 & 187 & -38 & -128 & 90 & 24 & -52 & 15 & 8 & -6 & 1 & 0 \\
&&&&&&&&&&&&&&&&&&&& \\
0 & -4 & 11 & 10 & -52 & 20 & 99 & -101 & -73 & 161 & -22 & -113 & 76 & 22 & -44 & 13 & 6 & -5 & 1 & 0 & 0 \\
&&&&&&&&&&&&&&&&&&&& \\
-1 & 4 & 2 & -23 & 11 & 48 & -52 & -39 & 83 & -10 & -59 & 35 & 13 & -19 & 3 & 3 & -1 & 0 & 0 & 0 & 0 \\
&&&&&&&&&&&&&&&&&&&& \\
0 & 1 & -4 & 1 & 13 & -14 & -11 & 25 & -4 & -16 & 11 & 2 & -4 & 1 & 0 & 0 & 0 & 0 & 0 & 0 & 0
\end{array}\right)
\end{equation*}
\begin{equation*}
H_{[2]}^{10_{122}}=\frac{1}{q^{22}A^{8}}
\left(\begin{array}{rrrrrrrrrrrrrrrrrrrrr}
0 & 0 & 0 & 0 & 0 & 0 & 0 & 1 & -3 & 0 & 9 & -8 & -8 & 15 & -2 & -10 & 7 & 1 & -3 & 1 & 0 \\
&&&&&&&&&&&&&&&&&&&& \\
0 & 0 & 0 & 0 & -1 & 2 & 4 & -13 & -2 & 34 & -19 & -42 & 49 & 16 & -48 & 13 & 18 & -13 & 0 & 3 & -1 \\
&&&&&&&&&&&&&&&&&&&& \\
0 & 0 & 1 & -4 & 3 & 15 & -29 & -14 & 78 & -32 & -95 & 97 & 40 & -99 & 22 & 42 & -27 & -3 & 8 & -2 & 0 \\
&&&&&&&&&&&&&&&&&&&& \\
0 & 1 & -5 & 3 & 20 & -34 & -24 & 93 & -28 & -120 & 103 & 59 & -118 & 19 & 58 & -39 & -6 & 16 & -5 & -2 & 1 \\
&&&&&&&&&&&&&&&&&&&& \\
1 & -5 & 3 & 19 & -32 & -17 & 84 & -24 & -95 & 90 & 46 & -90 & 22 & 41 & -30 & 1 & 9 & -3 & 0 & 0 & 0 \\
&&&&&&&&&&&&&&&&&&&& \\
-1 & 3 & 5 & -18 & -4 & 42 & -21 & -54 & 49 & 19 & -54 & 13 & 19 & -17 & 0 & 4 & -1 & 0 & 0 & 0 & 0 \\
&&&&&&&&&&&&&&&&&&&& \\
0 & 1 & -4 & 0 & 14 & -10 & -14 & 22 & 1 & -15 & 10 & 2 & -4 & 1 & 0 & 0 & 0 & 0 & 0 & 0 & 0
\end{array}\right)
\end{equation*}
\begin{equation*}
H_{[2]}^{10_{156}}=\frac{1}{q^{20}A^{10}}
\left(\begin{array}{rrrrrrrrrrrrrrrrrrr}
0 & 0 & 0 & 0 & 0 & 1 & -2 & 0 & 3 & -4 & 1 & 5 & -5 & -2 & 5 & -1 & -2 & 1 & 0 \\
&&&&&&&&&&&&&&&&&& \\
0 & 0 & -1 & 1 & 2 & -5 & 3 & 4 & -10 & 6 & 9 & -13 & -2 & 12 & -3 & -6 & 3 & 1 & -1 \\
&&&&&&&&&&&&&&&&&& \\
1 & -2 & 1 & 4 & -7 & 4 & 7 & -16 & 8 & 16 & -20 & -4 & 19 & -5 & -8 & 6 & 1 & -2 & 1 \\
&&&&&&&&&&&&&&&&&& \\
-1 & 1 & 1 & -5 & 4 & 4 & -14 & 6 & 14 & -16 & -5 & 13 & -2 & -6 & 2 & 1 & -1 & 0 & 0 \\
&&&&&&&&&&&&&&&&&& \\
0 & 1 & -1 & 1 & 2 & -6 & 3 & 7 & -8 & -2 & 6 & -1 & -2 & 1 & 0 & 0 & 0 & 0 & 0 \\
&&&&&&&&&&&&&&&&&& \\
0 & 0 & 0 & -1 & 0 & 2 & -1 & -1 & 1 & 0 & 0 & 0 & 0 & 0 & 0 & 0 & 0 & 0 & 0
\end{array}\right)
\end{equation*}
\begin{equation*}
H_{[2]}^{10_{158}}=\frac{1}{q^{18}A^{4}}
\left(\begin{array}{rrrrrrrrrrrrrrrrrrr}
0 & 0 & 0 & 0 & 0 & 0 & 0 & 0 & 0 & 0 & 0 & 0 & 0 & 1 & 0 & 0 & 0 & 0 & 0 \\
&&&&&&&&&&&&&&&&&& \\
0 & 0 & 0 & 0 & 0 & 0 & 0 & 0 & 0 & 1 & -2 & -1 & 4 & -1 & -3 & 1 & 1 & 0 & 0 \\
&&&&&&&&&&&&&&&&&& \\
0 & 0 & 0 & 0 & 0 & 1 & -3 & 1 & 7 & -12 & -2 & 18 & -9 & -11 & 9 & 1 & -4 & 0 & 0 \\
&&&&&&&&&&&&&&&&&& \\
0 & 0 & -1 & 1 & 2 & -8 & 5 & 14 & -24 & -2 & 34 & -16 & -19 & 20 & 3 & -9 & 3 & 2 & -1 \\
&&&&&&&&&&&&&&&&&& \\
1 & -2 & 2 & 5 & -11 & 5 & 21 & -29 & -8 & 43 & -17 & -25 & 24 & 1 & -11 & 5 & 1 & -2 & 1 \\
&&&&&&&&&&&&&&&&&& \\
-1 & 1 & 1 & -8 & 3 & 13 & -20 & -9 & 29 & -9 & -19 & 14 & 2 & -7 & 2 & 1 & -1 & 0 & 0 \\
&&&&&&&&&&&&&&&&&& \\
0 & 1 & -2 & 1 & 6 & -7 & -3 & 13 & -4 & -6 & 6 & 0 & -2 & 1 & 0 & 0 & 0 & 0 & 0
\end{array}\right)
\end{equation*}
\begin{equation*}
H_{[2]}^{10_{160}}=\frac{A^{4}}{q^{14}}
\left(\begin{array}{rrrrrrrrrrrrrrrrrrr}
0 & 0 & 0 & 0 & 0 & 0 & 0 & 0 & 0 & 0 & 0 & 0 & 0 & 0 & 1 & 0 & 0 & 0 & 0 \\
&&&&&&&&&&&&&&&&&& \\
0 & 0 & 0 & 0 & 0 & 0 & 0 & 0 & 0 & -1 & 1 & 1 & -3 & 1 & 2 & -2 & -1 & 0 & 0 \\
&&&&&&&&&&&&&&&&&& \\
0 & 0 & 0 & 0 & 0 & 1 & -1 & -2 & 1 & 2 & -2 & -3 & 3 & 2 & -1 & -1 & 0 & 2 & 0 \\
&&&&&&&&&&&&&&&&&& \\
0 & 0 & -1 & 1 & 3 & -2 & -3 & 3 & 4 & -4 & -4 & 5 & 1 & -3 & -1 & -1 & 1 & 0 & -1 \\
&&&&&&&&&&&&&&&&&& \\
1 & -2 & -1 & 5 & -2 & -6 & 4 & 4 & -5 & -2 & 4 & 1 & 0 & 0 & -2 & 3 & 1 & -2 & 1 \\
&&&&&&&&&&&&&&&&&& \\
-1 & 0 & 3 & -1 & -3 & 3 & 2 & -3 & -1 & 2 & 0 & 0 & 0 & -2 & 1 & 1 & -1 & 0 & 0 \\
&&&&&&&&&&&&&&&&&& \\
0 & 1 & -1 & -1 & 2 & 0 & -2 & 1 & 1 & -1 & 1 & 0 & -1 & 1 & 0 & 0 & 0 & 0 & 0
\end{array}\right)
\end{equation*}
\begin{equation*}
H_{[2]}^{10_{162}}=\frac{1}{q^{18}A^{12}}
\left(\begin{array}{rrrrrrrrrrrrrrrr}
0 & 0 & 0 & 0 & 0 & 0 & 0 & 3 & -1 & 0 & 5 & 0 & -2 & 3 & 1 & 0 \\
&&&&&&&&&&&&&&& \\
0 & 0 & 0 & 0 & -3 & 1 & 3 & -9 & -1 & 8 & -9 & -8 & 5 & 0 & -4 & -1 \\
&&&&&&&&&&&&&&& \\
0 & 0 & 2 & -5 & 3 & 9 & -15 & 1 & 22 & -13 & -11 & 16 & 3 & -6 & 2 & 1 \\
&&&&&&&&&&&&&&& \\
1 & 1 & -5 & 4 & 10 & -16 & -1 & 25 & -11 & -15 & 14 & 4 & -6 & 0 & 1 & 0 \\
&&&&&&&&&&&&&&& \\
0 & -3 & 2 & 6 & -12 & -3 & 17 & -8 & -12 & 8 & 2 & -3 & 0 & 0 & 0 & 0 \\
&&&&&&&&&&&&&&& \\
-1 & 1 & 2 & -5 & 0 & 7 & -3 & -4 & 3 & 1 & -1 & 0 & 0 & 0 & 0 & 0 \\
&&&&&&&&&&&&&&& \\
0 & 1 & -1 & 0 & 2 & -1 & -1 & 1 & 0 & 0 & 0 & 0 & 0 & 0 & 0 & 0
\end{array}\right)
\end{equation*}
\begin{equation*}
H_{[2]}^{10_{163}}=\frac{1}{q^{16}}
\left(\begin{array}{rrrrrrrrrrrrrrrrrrr}
0 & 0 & 0 & 0 & 0 & 0 & 0 & 0 & 0 & 0 & 0 & 0 & 0 & 1 & 0 & 0 & 0 & 0 & 0 \\
&&&&&&&&&&&&&&&&&& \\
0 & 0 & 0 & 0 & 0 & 0 & 0 & 0 & 0 & 1 & -1 & -4 & 2 & 2 & -4 & -1 & 1 & 0 & 0 \\
&&&&&&&&&&&&&&&&&& \\
0 & 0 & 0 & 0 & 0 & 1 & -4 & -1 & 14 & -8 & -16 & 22 & 6 & -17 & 7 & 5 & -3 & 0 & 0 \\
&&&&&&&&&&&&&&&&&& \\
0 & 0 & -1 & 2 & 4 & -11 & -2 & 27 & -16 & -32 & 35 & 8 & -34 & 11 & 11 & -11 & 1 & 3 & -1 \\
&&&&&&&&&&&&&&&&&& \\
1 & -3 & 1 & 9 & -15 & -7 & 36 & -13 & -37 & 40 & 12 & -36 & 16 & 11 & -15 & 5 & 2 & -3 & 1 \\
&&&&&&&&&&&&&&&&&& \\
-1 & 2 & 4 & -10 & -3 & 22 & -9 & -25 & 23 & 8 & -24 & 9 & 8 & -9 & 2 & 2 & -1 & 0 & 0 \\
&&&&&&&&&&&&&&&&&& \\
0 & 1 & -3 & -1 & 9 & -5 & -9 & 12 & 1 & -9 & 6 & 1 & -3 & 1 & 0 & 0 & 0 & 0 & 0
\end{array}\right)
\end{equation*}
\begin{equation*}
H_{[2]}^{10_{164}}=\frac{1}{q^{16}A^{8}}
\left(\begin{array}{rrrrrrrrrrrrrrrr}
0 & 0 & 0 & 0 & 0 & 0 & 0 & 3 & -3 & -4 & 8 & 1 & -6 & 1 & 1 & 0 \\
&&&&&&&&&&&&&&& \\
0 & 0 & 0 & 0 & -3 & 1 & 9 & -13 & -12 & 23 & 1 & -19 & 4 & 6 & -2 & -1 \\
&&&&&&&&&&&&&&& \\
0 & 0 & 2 & -6 & 1 & 21 & -18 & -22 & 41 & 6 & -31 & 10 & 12 & -6 & 0 & 1 \\
&&&&&&&&&&&&&&& \\
1 & 2 & -7 & 1 & 21 & -18 & -26 & 35 & 8 & -33 & 6 & 11 & -7 & -1 & 1 & 0 \\
&&&&&&&&&&&&&&& \\
0 & -5 & 2 & 14 & -14 & -16 & 26 & 5 & -20 & 6 & 6 & -3 & 0 & 0 & 0 & 0 \\
&&&&&&&&&&&&&&& \\
-1 & 2 & 5 & -8 & -6 & 13 & 0 & -9 & 3 & 2 & -1 & 0 & 0 & 0 & 0 & 0 \\
&&&&&&&&&&&&&&& \\
0 & 1 & -2 & -1 & 4 & -1 & -2 & 1 & 0 & 0 & 0 & 0 & 0 & 0 & 0 & 0
\end{array}\right)
\end{equation*}
\begin{equation*}
H_{[2]}^{10_{165}}=\frac{1}{q^{22}A^{16}}
\left(\begin{array}{rrrrrrrrrrrrrrrr}
0 & 0 & 0 & 0 & 0 & 0 & 0 & 3 & -3 & -4 & 7 & 1 & -5 & 1 & 1 & 0 \\
&&&&&&&&&&&&&&& \\
0 & 0 & 0 & 0 & -1 & -1 & 7 & -2 & -15 & 12 & 13 & -15 & -4 & 8 & 1 & -1 \\
&&&&&&&&&&&&&&& \\
0 & 0 & 1 & -2 & -2 & 12 & -2 & -21 & 14 & 19 & -18 & -9 & 9 & 2 & -3 & -1 \\
&&&&&&&&&&&&&&& \\
0 & 1 & -3 & -3 & 12 & -4 & -22 & 14 & 17 & -17 & -6 & 9 & 0 & -1 & 1 & 0 \\
&&&&&&&&&&&&&&& \\
1 & -3 & -1 & 11 & -5 & -16 & 14 & 11 & -13 & -2 & 5 & -1 & 0 & 0 & 0 & 0 \\
&&&&&&&&&&&&&&& \\
-1 & 1 & 5 & -4 & -8 & 9 & 5 & -8 & 0 & 2 & -1 & 0 & 0 & 0 & 0 & 0 \\
&&&&&&&&&&&&&&& \\
0 & 1 & -2 & -2 & 4 & 0 & -2 & 1 & 0 & 0 & 0 & 0 & 0 & 0 & 0 & 0
\end{array}\right)
\end{equation*}

\end{document}